\documentclass[a4paper,11pt]{article}
\pdfoutput=1 % if your are submitting a pdflatex \left(i.e. if you have
\usepackage{jcappub} % for details on the use of the package, please
\usepackage[T1]{fontenc} % if needed
\usepackage{caption}
\usepackage{subcaption}
\usepackage{graphicx}
\usepackage{amsthm}
\usepackage{comment}
\usepackage{hyperref}

\usepackage {amssymb}
\newcommand{\nc}{\newcommand}
\newcommand{\ba}{\begin{eqnarray}}
\newcommand{\ea}{\end{eqnarray}}

\newcommand{\calR}{{\cal{R}}}
\newcommand{\calP}{{\cal{P}}}

\def\bfk{{\bf k}}

\def\bfx{{\bf x}}

\nc{\cN}{ {\cal{N}} }

\title{  Inflation with Stochastic Boundary}

 \author{Amin Nassiri-Rad,}
 \author{Kosar Asadi,}
  \author{Hassan Firouzjahi}
 \affiliation{School of Astronomy, Institute for Research in Fundamental Sciences (IPM) \\ P.~O.~Box 19395-5531, Tehran, Iran}

\emailAdd{amin.nassiriraad@ipm.ir}
\emailAdd{k.asadi@ipm.ir}
\emailAdd{firouz@ipm.ir}
\abstract{  We study the Brownian motion of a field where there are boundaries in the inflationary field space.  Both the field and the boundary undergo Brownian motions with the amplitudes of the noises determined by the Hubble expansion rate of the corresponding dS spacetime. This setup mimics  models of inflation in which  curvature perturbation is induced from inhomogeneities generated at the surface of end of inflation. 
The cases of the drift dominated regime as well as  the diffusion dominated regime are studied in details. We calculate the first hitting probabilities  as well as the mean number of e-folds for the field  to hit either of the boundaries in the field space. The implications for models of inflation are reviewed. 
 }

\begin{document}

\maketitle

%%%%%%%%%%%%%%%%%%%%%%%%%%%%%%%%%%%%%%%%%%%%%%%%%%%%%%%%%%%%%%%%%%%%%%%%%%%%%%%%%%
\section{Introduction}

In the current  paradigm of early universe cosmology the large scale cosmological perturbations are believed  to be generated during inflation \cite{Guth:1980zm, Sato, Linde:1981mu, Albrecht}.  As a resolution for the flatness and the horizon problems,  the universe undergoes a period of accelerated expansion  in which small patches are stretched exponentially to encompass the current observable universe. In the simplest models, inflation is driven by a scalar field, the inflaton field, with a nearly flat potential. Classically, the inflaton field slowly rolls on top of its potential, while at the same time it undergoes quantum fluctuations.  These quantum perturbations are  generated continuously deep inside the Hubble horizon which are subsequently stretched to superhorizon scales as the universe expands exponentially.
The simplest models of inflation predict that these perturbations on large scales to be nearly scale invariant, nearly Gaussian and nearly adiabatic which are well consistent with 
cosmological observations \cite{Planck:2018vyg, Planck:2018jri}.

Stochastic inflation is an elegant method to study inflation and cosmological perturbations \cite{Vilenkin:1983xp, Starobinsky:1986fx}.  
It is an effective field theory approach for the dynamics of the long superhorizon perturbations which are  affected by small scale quantum perturbations. In this picture one decomposes the perturbations into the long and short perturbations. As the short modes are stretched beyond the horizon, they 
become classical and  affect the long mode perturbations. Using the stochastic formalism the effects of short modes on long modes are captured by random classical white noises with the amplitude $\frac{H}{2 \pi}$ in which $H$ is the Hubble expansion rate during inflation. Stochastic inflation has been used to study cosmological perturbations in slow-roll models \cite{ Rey:1986zk, Nakao:1988yi, Sasaki:1987gy, Nambu:1987ef, Nambu:1988je,  Kandrup:1988sc,  Nambu:1989uf, Mollerach:1990zf, Linde:1993xx, Starobinsky:1994bd, Kunze:2006tu, Prokopec:2007ak, Prokopec:2008gw, Tsamis:2005hd, Enqvist:2008kt, 
Finelli:2008zg, Finelli:2010sh, Garbrecht:2013coa, Garbrecht:2014dca, Burgess:2014eoa,  Burgess:2015ajz, Boyanovsky:2015tba,  Boyanovsky:2015jen, Pinol:2020cdp, Cruces:2018cvq, Cruces:2021iwq, Noorbala:2019kdd, Ahmadi:2022lsm},  ultra slow-roll setups \cite{Pattison:2019hef, Pattison:2021oen, Firouzjahi:2018vet, Firouzjahi:2020jrj}
and also in models involving gauge fields \cite{ Fujita:2017lfu, Fujita:2022fit, Fujita:2022fwc, Talebian:2019opf, Talebian:2020drj, Talebian:2021dfq, Talebian:2022jkb}.  Since in stochastic formalism one decomposes the perturbations into  long and short modes it provides  a natural setup to employ $\delta N$ formalism \cite{Fujita:2013cna, Fujita:2014tja, Vennin:2015hra, Vennin:2016wnk, Assadullahi:2016gkk,  Grain:2017dqa, Noorbala:2018zlv, Jackson:2022unc}. The standard $\delta N$ formalism \cite{Sasaki:1995aw, Sasaki:1998ug, Lyth:2004gb, Wands:2000dp, Lyth:2005fi, Abolhasani:2019cqw, Abolhasani:2018gyz} is based on the separate Universe approach where the superhorizon perturbations modify  the background expansion of the nearby patches (Universes). The $\delta N$ formalism is  a powerful tool to calculate the curvature perturbation correlations non-perturbatively, i.e. to all orders in perturbations.  Using the stochastic $\delta N$ formalism the stochastic corrections in curvature perturbation power spectrum and bispectrum and the consequences for primordial black hole formation can be studied.

In the past studies involving the formalism of stochastic inflation, the stochasticity was 
imprinted from the dynamics of the rolling inflaton or vector fields during inflation. In these works, the end of inflation is a fixed point in field space so there is no curvature perturbations generated from the surface of end inflation. In these scenarios, one only needs to follow the stochastic dynamics of the fields  during inflation to read off the amplitude of curvature perturbations generated during inflation on superhorizon scales. However, there are interesting examples where curvature perturbations can be induced  from inhomogeneities generated at the surface of end of inflation \cite{Lyth:2005qk, Sasaki:2008uc, Naruko:2008sq, Dvali:2003em, Abolhasani:2013bpa, Hooshangi:2022lao}. These are multiple field scenarios in which inflation is terminated on a surface in a field space. As the surface of end of inflation is modulated by inhomogeneities induced from the multiple fields, then curvature perturbations are also generated at the surface of end of inflation. In the language of stochastic inflation, we are dealing with a setup where the boundary itself undergoes stochastic motion. Our goal in this work is to extend the analysis of stochastic inflation to the above mentioned picture where we have additional source of stochasticity from the boundaries in the field space. With this motivation in mind, we study various physical and mathematical questions related to dS backgrounds with random fields and stochastic boundaries.

The paper is organized as follows. In section \ref{review} we briefly review the setup of stochastic inflation and a simple example of inflation where inhomogeneities are 
generated at the surface of end of inflation. In section  \ref{driftdominate} we study the case in which the classical drift of the field is dominant compared to the Brownian motions of the field and the boundary. In section \ref{diffdominate} we extend these analysis 
to the case where the system is diffusion dominated while the Brownian boundary is confined to move within two fixed barriers. In section \ref{Uniform2} we study 
the setup where the stochastic motion of the boundary has a uniform distribution 
followed by Summary and Discussions in section \ref{summary}. Some technical analysis are relegated into Appendices \ref{proof} and \ref{prob}.

%%%%%%%%%%%%%%%%%%%%%%%%%%%%%%%%%%%%%%%%%
\section{Review and Motivation }
\label{review}

In this section we briefly review the formalism of stochastic inflation and present our motivation in studying the models containing boundaries which  undergo Brownian motion. 

Consider a single field inflation model containing the inflaton field $\phi$ which slowly rolls under the potential $V(\phi)$.  We split $\phi$ and its conjugate momentum 
$v =\dot \phi$ into the short and long wavelengths, 
\begin{equation}
\label{3}
\begin{split}
\phi\left(\bfx,t\right)=\phi_l\left(\bfx,t\right)+\sqrt{\hbar}\phi_s\left(\bfx,t\right),
\end{split}
\end{equation}
\begin{equation}
\label{4}
\begin{split}
v\left(\bfx,t\right)=v_l\left(\bfx,t\right)+\sqrt{\hbar}v_s\left(\bfx,t\right),
\end{split}
\end{equation}
in which the subscripts  $l$ and $s$ denote the long modes and short modes respectively. The factor $\sqrt \hbar$ has been inserted explicitly for the short modes  above  to specify the quantum nature of the short modes. 

Going to the Fourier space,  the short modes satisfy the following decompositions, 
\begin{equation}
\label{5}
    \phi_s\left(\bfx,t\right)=\int\frac{d^3k}{\left(2\pi\right)^3}\theta\left(k-\varepsilon aH\right)\phi_\bfk\left(t\right)e^{ik.\bfx},
\end{equation}
and
\begin{equation}
\label{6}
    v_s\left(\bfx,t\right)=\int\frac{d^3k}{\left(2\pi\right)^{3}}\theta\left(k-\varepsilon aH\right)\dot\phi_\bfk\left(t\right)e^{ik.\bfx}.
\end{equation}
Here the dimensionless parameter $ \varepsilon$ is a small number $\varepsilon \ll 1$ which is introduced to separate the large and small scales. Furthermore, $a(t)$ is the scale factor and $H=\dot a/a$ is the Hubble expansion rate during inflation.
The operator $\phi_{\bfk }(t)$ is given by  $\phi_\bfk=a_\bfk\varphi_k+a^\dagger_{-\bfk}\varphi_{-k}^*$ in which $\varphi_k$ is the positive frequency mode function satisfying the scalar field equation of motion 
(the Klein-Gordon equation in the presence of gravity) while $a_\bfk$ and $a^\dagger_{\bfk}$ are the usual annihilation  and creation operators. 

By expanding the scalar field Klein-Gordon equation  around $\phi_l$ and $v_l$ up to first order in $\sqrt{\hbar}$ one obtains the following equations of motion for $\phi_l$ and $v_l$ \cite{Nakao:1988yi, Sasaki:1987gy}
\ba
    \label{7}
    \dot \varphi_l &=&v_l+\sqrt{\hbar}\sigma,\\
    \label{8}
    \dot{v}_l &=&-3Hv_l+\frac{1}{a^2}\nabla^2\varphi_l-V'\left(\varphi\right)+\sqrt{\hbar}\tau,
\ea
in which the noises  $\sigma$ and $\tau$, appearing as the source terms in 
Eqs. (\ref{7}) and (\ref{8}), 
are given by
\begin{equation}
    \label{9}
    \sigma\left(\bfx,t\right)= \varepsilon aH^2\int\frac{d^3 \bfk}{\left(2\pi\right)^3}\delta\left(k-  \varepsilon aH\right)\phi_\bfk\left(t\right)e^{i \bfk \cdot \bfx},
\end{equation}
\begin{equation}
    \label{10}
    \tau\left(\bfx,t\right)= \varepsilon aH^2\int\frac{d^3 \bfk}{\left(2\pi\right)^3}\delta\left(k-  \varepsilon aH\right)\dot\phi_\bfk\left(t\right)e^{i \bfk \cdot \bfx}.
\end{equation}

Starting with the Bunch-Davies initial condition $|0 \rangle$, the commutation of the field and its momentum is given by  \cite{Nakao:1988yi, Sasaki:1987gy} 
\ba
    \left[\sigma\left( \bfx_1, t_1\right),\tau\left(\bfx_2, t_2\right)\right] &= & i \varepsilon^3\frac{H^4}{4\pi^2}j_0\big( \varepsilon aH | \bfx_1-\bfx_2|\big)\delta\left(t_1-t_2\right) \, ,
\ea
in which $j_0$ is the zeroth order Bessel function. As it can be seen from the above commutation relation,  the quantum nature of $\sigma$ and $\tau$ disappears if we choose $\varepsilon$ small enough. Furthermore, calculating the auto-correlation of the noises, one can show that $\tau=O\left( \varepsilon^2\right)$ while 
\begin{equation}
    \label{stoch}
     \left<0\left|\sigma\left(\bfx, t_1\right)\sigma\left(\bfx, t_2\right)\right|0\right>\approx\frac{H^3}{4\pi^2}\delta\left(t_1-t_2\right)=\frac{H^4}{4\pi^2}\delta\left(N_1-N_2\right)\, ,
\end{equation}
where $N$ is the number of e-fold related to cosmic time via $d N= H d t$. 
In the slow-roll limit which we are interested in, we can simply set $N= H t$.

On the super horizon limit $k \ll aH$ we can neglect the second order spatial derivative  in Eq. (\ref{8}). In addition, setting  $\tau \rightarrow 0$ (note that $\tau=O\left( \varepsilon^2\right)$) and dropping  the subscript $l$  for convenience,  from the combination of Eqs. (\ref{7}) and (\ref{8}) we obtain the following Langevin equation for the  coarse grained long mode 
\begin{equation}
\label{Langevin}
  \frac{d\phi(N)}{d N}=-\frac{V_{,\phi}}{3H^2}+\frac{H}{2\pi}\xi(N),
\end{equation}
where the new noise $\xi(N)$ is related to the original noise via 
$\sigma\equiv\frac{H}{2\pi}\xi$ so $\xi$ is a normalized classical white noise satisfying 
\ba
\label{xi-noise}
\big \langle \xi\left(N\right)\big \rangle = 0 \, , \quad \quad 
\big \langle \xi\left(N\right)\xi\left(N'\right)\big \rangle =\delta\left(N-N'\right) \, .
\ea 

Associated to the normalized noise $\xi(N)$ we can define a Wiener process 
$W(\cN)$ which is defined via $W(\cN)\equiv \int_0^{\cN} \xi(N) d N$ which satisfies the following conditions
\ba
\label{Wiener}
\langle W(\cN) \rangle =0 \, , \quad \quad  \langle W(\cN)^2 \rangle = \cN \, .
\ea

\subsection{Motivation for inflation with stochastic boundary}

Having obtained the Langevin equation  in Eq. (\ref{Langevin}) one can look at various of its applications. For example, one can calculate the average number of e-fold for inflation $\langle \mathcal{N} \rangle$ and the curvature perturbation power spectrum and their stochastic corrections \cite{Vennin:2015hra, Vennin:2016wnk, Assadullahi:2016gkk}. In deriving Eq. (\ref{Langevin}) we have assumed that only one field is responsible for curvature perturbation so stochasticity is generated purely from 
$\delta \phi$ perturbations during inflation. 
In particular, we assume that inflation is terminated at a specific point in field space where $\phi= \phi_e$. As the point of end of inflation is fixed, $\delta \phi_e=0$, then curvature perturbations are exclusively generated during inflation. For example, going to flat slicing, the comoving curvature perturbation is given by $\calR = -\frac{H}{\dot \phi} \delta \phi$ where $\delta \phi$ is calculated at the time of horizon crossing when $k= a H$. 

As a simple extension to multiple field setup, now consider the model containing two fields, the inflaton $\phi$ and the spectator field $\sigma$. The field $\sigma$ is massless and it does not contribute to potential and the background expansion.  However, it affects the surface of end of inflation.  As the field $\sigma$ modulates the surface of end of inflation, its perturbations generate additional contribution into curvature perturbations so the total curvature perturbation now is given by
\ba
\label{R-tot}
\calR = -\frac{H}{\dot \phi} \delta \phi + \calR_e
\ea
in which $\calR_e$ represents the curvature perturbations induced from the surface of end of inflation. 

As a specific example, consider the model studied in \cite{Lyth:2005qk} 
(see also \cite{Sasaki:2008uc}) in which the surface of end of inflation is determined 
by the ellipse
\ba
\label{ellipse}
\lambda_\phi \phi^2 + \lambda_\sigma \sigma^2 = M^2 \, ,
\ea
in which $M$ is a fixed mass scale and $\lambda_\phi$ and $\lambda_\sigma$ are
two coupling constants. 
In the absence of the spectator field $\sigma$, i.e. when $\lambda_\sigma=0$, inflation ends at a fixed point $\phi_e =\sqrt{M/\lambda_\phi}$. However, in the presence of the spectator field, there can be additional perturbations generated at the surface of end of inflation via
\ba
\label{delta-ratio}
\delta \phi_e = -\frac{\lambda_\sigma \sigma_e}{\lambda_\phi \phi_e} \delta \sigma \, .
\ea
So the total curvature perturbation from Eq. (\ref{R-tot}) is obtained to be
\ba
\label{R-tot2}
\calR = -\frac{H}{\dot \phi}  \Big( \delta \phi - \frac{\lambda_\sigma \sigma_e}{\lambda_\phi \phi_e} \delta \sigma \Big) \, .
\ea
As the perturbations $\delta \phi$ and $\delta \sigma$ are uncorrelated, each with the amplitude $H/2 \pi$, the curvature perturbation power spectrum $\calP_\calR$ is 
obtained to be
\ba
\label{power-total}
\calP_\calR =  \frac{H^2}{8 \pi^2 M_P^2 \epsilon_e} \Big[1+ \big( \frac{\lambda_\sigma \sigma_e}{\lambda_\phi \phi_e}\big)^2
\Big] \, ,
\ea
in which $\epsilon = M_P^2(V_{,\phi}/V)^2/ 2 $ is the slow-roll parameter associated with the rolling of $\phi$ and $M_P$ is the reduced Planck mass. The second term in the big bracket above represents the corrections induced from the quantum fluctuations generated at the surface of end of inflation. 

Motivated by the above example, our goal in this work is to study the setups where boundaries undergo stochastic motion. To be specific, we study a one dimensional 
stochastic process, denoted by the field $\phi(N)$,  
surrounded by two boundaries. The boundary at the left is denoted by $\phi_-$ while the right boundary is denoted by $\phi_+$.  As a warm-up exercise, first we 
consider the example described above and study the system using  the formalism of stochastic $\delta N$. After that we consider more complicated case where the system is diffusion dominated so the field $\phi$ and the boundary $\phi_+$ move stochastically under their Brownian motions.

%%%%%%%%%%%%%%%%%%%%%%%%%%%%%%%%%%%%%%%%%

\section{Stochastic Boundary in Drift Dominated Regime}
\label{driftdominate}

In this section, as a warm-up exercise,  we study the example presented in previous section using the formalism of stochastic inflation. It is a one dimensional drift dominated setup where the dynamics of the field $\phi(N)$ is governed by the Langevin equation (\ref{Langevin}). In the limit of slow-roll motion where one can neglect the evolution of the slow-roll parameter $\epsilon$, the Langevin equation (\ref{Langevin}) can be integrated yielding 
\begin{equation}
\label{phi-eq}
    \phi(N)=\phi_0+ C N+  A\,  W(N),
\end{equation}
in which for our case of interest $C \equiv \sqrt{2 \epsilon} M_P$ and $A \equiv \frac{H}{2 \pi}$. Also note that $W(N)$ is a Wiener process defined in Eq. (\ref{Wiener}).   Here  we work in the drift dominated regime where the dynamics of the field is governed by the classical drift term, corresponding to $A \ll C$. This is equivalent to the condition $\calP_\calR \ll1$ which  is assumed to be the case here.  

In principle there can be two boundaries, the left boundary denoted by $\phi_-(N)$ and the right boundary, $\phi_+(N)$. In the example presented in previous section the left boundary is placed at infinity. Physically, this corresponds to setting the UV scale to a very high value, say the scale of quantum gravity, so the field never explore those regions. Correspondingly, here we also assume that the left boundary is pushed to infinity, $\phi_-\rightarrow  -\infty$, so we have only the right boundary  $\phi_+$ which undergoes Brownian motion. 

 The equation of motion for the right boundary  in this case is given by
\begin{equation}
\label{sigma-eq}
    \phi_{+}(N)=\phi_{+}^{(0)}+ B\,  W_+(N),
\end{equation}
where $\phi_{+}^{(0)}$ represents the initial position of the right boundary and 
$B$ is  a constant, determining the amplitude of its Brownian motion. Also 
note that the stochastic natures of the field and the boundary are different so the two Wiener processes $W(N)$ and $W_+(N)$ are uncorrelated, $\langle W(N)  W_+(N) \rangle =0$.  

Our goal is to solve the system of equations (\ref{phi-eq}) and (\ref{sigma-eq}) jointly 
to obtain observable quantities like $\langle \cN \rangle$ and $\calP_\calR$ using the stochastic $\delta N$ formalism.  A simple approach is to change the frame and go to the reference frame of the right boundary. In this case, the boundary is fixed but its stochastic motion is transferred to the field $\phi(N)$, so the position of field in this frame is given by 
\begin{equation}
\label{phi-eq2}
    \phi(N)=\phi_0 + C N + A \, W(N)-B\,  W_+(N).
\end{equation}

As $W(N)$ and $W_+(N)$  are two random Gaussian processes which 
evolve independently of each other,  their combination represents  a new random Gaussian process as follows,
\begin{equation}
   A \, W(N)-B \, W_+(N) \rightarrow \sqrt{A^2+B^2} \, W_n(N) \, ,
\end{equation}
in which $W_n(N)$ is the new Wiener process. 
As before,  to preserve the drift dominated regime,  we assume that the constraint 
$C\gg\sqrt{A^2+B^2}$ holds as well.

Let us denote the time when the field hits the boundary by $\cN$. Note that while $N$ is the clock which is deterministic, the quantity $\cN$ is a stochastic variable. 
Our goal is to calculate  $\langle  \cN \rangle$ and 
$\delta \cN^2 \equiv \langle  \cN^2 \rangle- \langle  \cN \rangle^2$. 
To calculate these correlations one can use the first boundary crossing approach
\cite{Vennin:2015hra}. As we have two boundaries in field space, the field hits either of boundaries with different probabilities. We denote $p_-$ as the probability of the field hitting $\phi_-$ ($\phi_+$) {\it first} before hitting the other boundary at $\phi_+$ ($\phi_-$).   Note that by construction $p_- + p_+ =1$. 
In our current setup we have pushed the left boundary to infinity, $\phi_{-} \rightarrow -\infty$. As demonstrated in the Appendix C of \cite{Talebian:2022jkb} in the limit 
that  $\phi_{-} \rightarrow -\infty$ one obtains  $p_- \phi_{-} =0$ which will be used in our analysis below. 

Taking the stochastic average of both sides of
Eq. (\ref{phi-eq2}) we obtain  
 \begin{equation}
 \label{N-eq1}
    \big\langle {\phi(\cN) -\phi_0}\big \rangle=C \langle \mathcal{N} \rangle \, ,
 \end{equation}
 where we have used the fact that $\langle  W_n(\cN) \rangle =0$. On the other hand, from the definition of first hitting probability we have 
 \begin{equation}
 \label{av-phi-eq1}
    \big\langle {\phi-\phi_0}\big \rangle=p_{+}(\phi_{+}^{(0)}-\phi_{0})+p_{-}(\phi_{-}-\phi_{0})=\phi_{+}^{(0)}-\phi_{0} \, ,
 \end{equation}
where the condition  $p_- \phi_{-} =0$ has been used \cite{Talebian:2022jkb}. Combining Eqs. (\ref{N-eq1}) and (\ref{av-phi-eq1}) we obtain 
\begin{equation}
\label{av-N}
    \big<\mathcal{N}\big>=\frac{\phi_{+}^{(0)}-\phi_0}{C},
\end{equation}
which is consistent with what one may expect from the classical evolution. 

To calculate $\langle  \cN^2 \rangle$ we take the average of the square of Eq. (\ref{phi-eq2}) as follows \cite{Talebian:2022jkb},
\begin{equation}
\label{4-16a}
\begin{split}
    &\big\langle {(\phi-\phi_0)^2}\big\rangle=(\phi_{+}^{(0)}-\phi_{0})^2\\&
    \hspace{1.9cm}=C^2 \big \langle \mathcal{N}^{2} \big\rangle + 2 C \sqrt{A^2+B^2} \,\,\big\langle  W(\mathcal{N}) \mathcal{N}\big \rangle  + (A^2+B^2)\big\langle {W (\mathcal{N})}^2\big\rangle \, .
\end{split}
 \end{equation}
To proceed, one should calculate $\big\langle W(\mathcal{N}) \mathcal{N}\big\rangle$. For this purpose, from Eq. (\ref{phi-eq2}),  we note that 
 \ba
  \big \langle  W(\mathcal{N}) \mathcal{N}\big \rangle = \frac{1}{C}\Big\langle W(\mathcal{N})\Big( {\phi-\phi_{0}-\sqrt{A^2+B^2} W(\mathcal{N})}  \Big)
  \Big  \rangle  \,. 
   \ea
The right hand side of the above equation is decomposed into 
\ba
    \frac{p_{+}}{C}\big({\phi_{+}^{(0)}-\phi_{0}}\big)\big\langle W(\mathcal{N}) | \phi=\phi_{+}^{(0)}\big\rangle + \frac{p_{-}}{C}\big( \phi_{-}-\phi_{0} \big)\big\langle W(\mathcal{N}) | \phi=\phi_{-}\big\rangle - \frac{\sqrt{A^2+B^2} } {C}\big\langle\mathcal{N}\big\rangle  .
\ea
Setting $p_{-}\phi_{-}=0$ and $\big< W(\mathcal{N}) | \phi=\phi_{+}^{(0)}\big>=\big< W(\mathcal{N})\big>=0$ the first two terms above vanish. Correspondingly, we obtain
\ba
\label{WN}
 \langle  W(\mathcal{N}) \mathcal{N} \rangle = - \frac{\sqrt{A^2+B^2} }{C}\big\langle\mathcal{N}\big\rangle \, .
\ea
Plugging Eq. (\ref{WN}) in Eq. (\ref{4-16a}) and noting that 
$\big\langle {W (\mathcal{N})}^2\big\rangle = \langle \cN \rangle$ we obtain 
\begin{equation}
\begin{split}
   (\phi_{+}^{(0)}-\phi_{0})^2 = C^2 \big\langle \mathcal{N}^{2} \big\rangle- 
    (A^2+B^2) \big\langle \cN\big\rangle \, .
\end{split}
 \end{equation}
 Combining this result with the expression for $\langle \cN \rangle$ in 
 Eq. (\ref{av-N}) we obtain 
\begin{equation}
    \big\langle \mathcal{N}^2\big\rangle=  \langle \cN \rangle^2 +
     \frac{A^2+B^2}{C^2}  \langle \cN \rangle \, ,
\end{equation}
yielding, 
\ba
\delta \cN^2 =  \frac{A^2+B^2}{C^2}  \langle \cN \rangle  \, .
\ea
The power spectrum of curvature perturbation is related to the variance via \cite{Vennin:2015hra} 
\ba
\calP_\calR = \frac{d \,  \delta \cN^2}{d  \langle \cN \rangle}\, ,
\ea
yielding 
\begin{equation}
    \mathcal{P}_\calR=\frac{A^2+B^2}{C^2} \, .
\end{equation}
In our setup of slow-roll inflation with $A=\frac{H}{2\pi}$ and $C=\sqrt{2\epsilon} M_P$ one obtains,
\begin{equation}
\label{power2}
    \mathcal{P}_\calR= \frac{H^2}{8 \pi^2 M_P^2 \epsilon} 
    \Big(1+\frac{B^2}{A^2} \Big)  \, .
   % \mathcal{P}_{\zeta_0}+\frac{B^2}{2\epsilon}.
\end{equation}

Now we apply the above formula to our  particular example of inhomogeneities generated from the surface of end inflation with the boundary given in Eq. (\ref{ellipse}). In this example the quantity  $B$ measures the amplitude of the fluctuations induced by the spectator field at the end of inflation. In other words, the stochastic behaviour of the boundary is played by the quantum fluctuations of the spectator field $\sigma$. 
Noting that the amplitude of both fluctuations $\delta \phi$ and $\delta \sigma$ on the initial flat 
hypersurface is equal to $H/2\pi$, the ratio 
$\frac{B}{A}$  is determined from Eq. (\ref{delta-ratio}) to be  
$\frac{B}{A}= -\frac{\lambda_\sigma \sigma_e}{\lambda_\phi \phi_e}$. Plugging this value in Eq. (\ref{power2}) yields the result  Eq. (\ref{power-total}) for the total power spectrum.

%%%%%%%%%%%%%%%%%%%%%%%%%%%%%%%%%%%%%%%%%%%%%%%%%%%%%%%%%%%%%%%%%%%%%%%%%%%%%%%%%%
\section{Diffusion Dominated Regime}
\label{diffdominate}

Here we study our main case of interest where the system is diffusion dominated, i.e. the field $\phi$ is under a Brownian motion with the amplitude $\frac{H}{2\pi}$. We may allow for a subleading drift term, but the leading effects in the dynamics  of the system are controlled by the diffusion term in the corresponding Langevin equation. 
Physically, this corresponds to the case where $\calP_\calR >1$. Obviously this can not happen in the context of single field slow-roll inflation (at least on CMB scales). However, this can happen in the general landscape of inflationary cosmology such as in the context of eternal inflation. Also, we may consider some non-trivial setups of inflation in which  locally on a short period of inflation  
$\calP_\calR$ is amplified to order unity such as in the mechanism of primordial black hole formation in USR setup \cite{Pattison:2019hef, Pattison:2021oen}. Therefore,  our current setup of diffusion dominated regime may be relevant for those setups as well. 

We consider two absorbing boundaries in field space, the left boundary $\phi_-$ and the right boundary $\phi_+$. In principle both boundaries can undergo Brownian motion. However, to simplify the situation we assume that only the right boundary undergoes Brownian motion while the left boundary is held fixed. 
As we shall see below, the analysis even in this simplified case is complicated and rich that worths this approximation. This approximation may be well-motivated physically. As we argued before, one can view the left boundary to be located 
in the  UV region which may not be explored in our perturbative approach. On the other hand, the right boundary may be viewed as the surface of reheating where many fields can contribute to its modulation as for example in  the case of the boundary given in Eq. (\ref{ellipse}). Having said this, after developing the formalism here for the case where only the right boundary is stochastic, one can extend the current formalism to more complicated setup where both boundaries move stochastically. 

With the above discussions in mind the mathematical description of the evolution of the field and the right boundary is given as follows,\footnote{We have denoted the Wiener process of $\phi_+$ by $\Tilde{W}_+(N)$ to emphasis that it is confined between two reflective barriers. }
\begin{equation}
\label{langevin-phi-plus}
    \phi_{+}(N)=\phi_{+}^{(0)}+D\,  \Tilde{W}_+(N),
\end{equation}
\begin{equation}
\label{langevinphi}
    \phi(N)=\phi_0+W(N),
\end{equation}
where, as before, $\phi_{+}(N)$  and $\phi(N)$ describe the evolution of the right boundary and the field respectively while  $\phi_{+}^{(0)}$  and $\phi_0$ represent their initial values.  Here, we have scaled the fields in the unit of  $\frac{H}{2\pi}$ so if our primary  field is $\Phi(N)$ then the new rescaled field is  defined via $\phi(N) \equiv \frac{\Phi(N)}{{H}/{2\pi}}$. Finally,  $D$ represents the amplitude of the diffusion associated to $\phi_+$ while the  the diffusion amplitude associated to the field is set to unity (in the unit of  $\frac{H}{2\pi}$). The two Wiener processes  $\Tilde{W}_+(N)$ and $W(N)$ are independent so  $  \big \langle W(N)\Tilde{W}_+(N)\big \rangle=0$.  

\begin{figure} 
    \centering
    \includegraphics[scale=0.63]{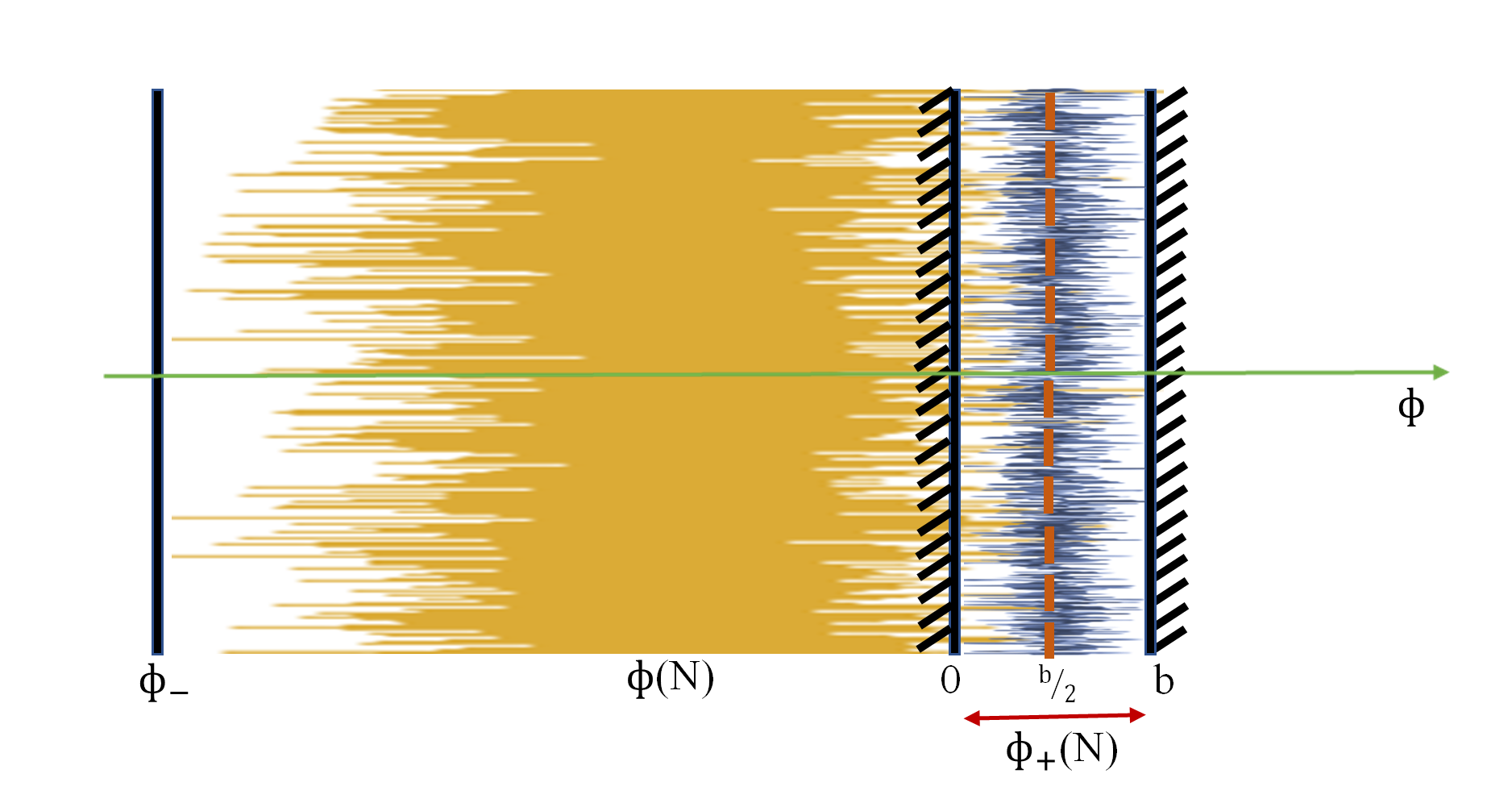}
    \caption{A schematic view of the setup in the case of diffusion dominated regime. The stochastic behaviour of the field  $\phi(N)$ is shown by an orange colour noise while that of the right boundary is denoted by  a blue colour noise. The field $\phi$ is restricted to move in the interval $[\phi_-, \phi_+]$ while the right boundary 
 is restricted to move between two reflective barriers which are fixed at $0$ and $b$. For a better understanding, the plot is presented as two-dimensional  in which the vertical axis represents the direction of time. Therefore, each slice of the diagram along the $\phi$ axis indicate the positions of $\phi$ and $\phi_+$.}
    \label{diagram}
\end{figure}

A schematic view of the setup is presented in Fig. \ref{diagram}. For a pictorial understanding, the stochastic nature of the field $\phi(N)$ is shown with the orange colour noise while the Brownian motion of the boundary is represented by a blue colour noise. It is worth mentioning that for a better understanding of the setup, this diagram is two-dimensional in which the vertical axis denotes the arrow of time so  
each slice of the diagram along the $\phi$ axis indicates the values of the field and the boundary.

As we mentioned before, the left boundary is fixed at a finite distance in one dimensional field space at the position $\phi_-$. The field $\phi$ moves stochastically in the range  $[\phi_-, \phi_+]$ while the right boundary $\phi_+$ 
undergoes stochastic motions as well. While the field $\phi$ moves stochastically in the range  $[\phi_-, \phi_+]$ the boundary $\phi_+(N)$ may hit the left boundary in a jump. This brings additional complexity in our analysis where we are primarily interested in the first hitting probability of the field to the boundaries before the boundaries themselves collide with each other. To bypass this difficulty we impose one more constraint that the stochastic movement of the right boundary is limited between two barriers separated by a distance  $b$ such that  the left boundary is always outside this range (see Fig. \ref{diagram}).  We choose our coordinate system (without loss of generality)  such that the two barriers are located at $0$ and 
$b$ so the stochastic motion of the right boundary is in the interval $[0, b]$. Within this description $\phi_- <0$, so the right boundary never hit the left boundary.
Moreover, since we do not want the boundary $\phi_+$ to be absorbed by the two barriers, we choose both of the barriers to be reflective. To prevent confusion, we adopt the terminology  ``boundary'' for the two end points $\phi_-$ and $\phi_+$
where the field can move while the two fixed points at $0$
and $b$ where $\phi_+$ is confined to is referred to as the  ``barriers''.

%%%%%%%%%%%%%%%%%%%%%%%%%%%%%%%%%%%%%%%%%
%%%%%%%%%%%%%%%%%%%%%%%%%%%%%%%%%%%%%%%%%

With the above discussions in mind,  the time dependent probability density function (PDF) associated to the Brownian movement of the right boundary, $f_+$,
 is described by the Fokker-Planck equation as follows,
\begin{equation}\label{fokkerbarrier1}
    \frac{\partial f_{+}(\phi_{+},N)}{\partial N}=\frac{D^2}{2}\frac{\partial^2 f_{+}(\phi_{+},N)}{\partial \phi_+^2}.
\end{equation}
As $\phi_+$ is limited in the interval $[0, b]$ with reflective barriers,  $f_{+}(\phi_+,N)$ satisfies the following Neumann boundary conditions,
\begin{equation}\label{BC}
  \frac{\partial f_{+}(\phi_{+},N)}{\partial \phi_{+}}\big|_{\phi_{+}=0}=\frac{\partial f_{+}(\phi_{+},N)}{\partial \phi_{+}}\big|_{\phi_{+}=b}=0 \, ,
\end{equation}
with the following initial condition 
\begin{equation}\label{delta}
  f_{+}(\phi_{+},N=N_0)=\delta(\phi_{+}-\phi_{+}^{(0)}),
\end{equation}
in which $N_0$ is the initial time when the boundary starts to evolve. Without loss of generality we may set $N_0=0$. 

Using the  method of the separation of variables
the general solution to Eq. \eqref{fokkerbarrier1} is given as 

\begin{equation}
f_k(x,t)=A_k \sin(k x)+B_k \cos(k x).
\end{equation}
With the boundary conditions Eqs. \eqref{BC} one can easily deduce that $k=\frac{m\pi}{b}$ where $m$ is a  non-negative integer numbers and $A_k=0$. Then the solution satisfying Eq. \eqref{delta} is given by
\begin{equation}
\label{fp}
    f_{+}(\phi_{+},N)=\frac{1}{b}+\frac{2}{b}\sum_{m=1}^{\infty}\cos\left(\frac{m\pi }{b}\phi_{+}^{(0)}\right)\cos\left(\frac{m\pi }{b}\phi_{+}\right)e^{-\frac{m^2\pi^2 D^2}{2 b^2}N}.
\end{equation}
It can be easily checked that the above PDF is normalized in $[0,b]$. Results similar to  Eq. \eqref{fp} are obtained in \cite{finite:region}. As we see, if $D\rightarrow0$ the PDF of the Brownian boundary reduces to $\delta (\phi_+ - \phi_+^{(0)})$ while in the limit $D\rightarrow\infty$ the PDF tends to forget its initial condition and 
$f_{+}(\phi_{+},N) \rightarrow \frac{1}{b}$.    Moreover, we note that as the width separating the two  barriers, $b$, goes to zero the above distribution function reduces to $\delta (\phi_+)$ as there is no room for the boundary to fluctuate.

Moreover, one can check that the above PDF enjoys the following symmetry. If the boundary starts at $\phi_+^{(0)}$, the probability of finding the boundary in the interval $0<\phi_+<\alpha$ at time $t$ is equal to finding the boundary in the interval $b-\alpha<\phi_+<b$ if it starts at $b-\phi_+^{(0)}$.
One can verify this as follows:
%\begin{equation}
%\begin{split}
\ba\label{symmetry}
&&\int_{b-\alpha}^b\left[\frac{1}{b}+\frac{2}{b}\sum_{m=1}^{\infty}\cos\left(\frac{m\pi }{b}(b-\phi_{+}^{(0)})\right)\cos\left(\frac{m\pi }{b}\phi_{+}\right)e^{-\frac{m^2\pi^2 D^2}{2 b^2}N}\right] d\phi_+ \nonumber\\
&=&\int_{0}^\alpha\left[\frac{1}{b}+\frac{2}{b}\sum_{m=1}^{\infty}(-1)^m\cos\left(\frac{m\pi }{b}\phi_{+}^{(0)}\right)\cos\left(\frac{m\pi }{b}(b-\phi_{+})\right)e^{-\frac{m^2\pi^2 D^2}{2 b^2}N}\right]d\phi_+\nonumber\\
&=&\int_{0}^\alpha\left[\frac{1}{b}+\frac{2}{b}\sum_{m=1}^{\infty}\cos\left(\frac{m\pi }{b}\phi_{+}^{(0)}\right)\cos\left(\frac{m\pi }{b}\phi_+\right)e^{-\frac{m^2\pi^2 D^2}{2 b^2}N}\right]\,d\phi_+\,.
\ea
%\end{split}
%\end{equation}

It is also instructive to study the behaviour of $f_+(\phi_+,N)$ when $b \rightarrow \infty$ so the right boundary can move arbitrarily large distance along the positive axis direction.  To this end we can replace the summation in Eq. (\ref{fp})
into an integration using the following relation 
\begin{equation}\label{seriesint}
  \lim_{b\rightarrow\infty}\sum_{i=1}^\infty\frac{1}{b}G\big(\frac{i}{b}\big)=
  \int^1_0 G(x)\, dx \, ,
\end{equation}
obtaining, 
\begin{equation}\label{fpint}
\begin{split}
 b_{+}(\phi_{+},N)=
   %\\&=\lim_{b\rightarrow\infty} \left[\frac{1}{b-\phi_-}+\frac{2}{b-\phi_-}\sum_{m=1}^{\infty}\left[\cos\left(\frac{m\pi (\phi_{+}^{(0)}-\phi_-)}{b-\phi_-}\right)\cos\left(\frac{m\pi (\phi_{+}-\phi_-)}{b-\phi_-}\right)e^{-\frac{m^2\pi^2 }{2 (b-\phi_-)^2}N}\right]\right]\\&=
   2\int^\infty_0\cos\left(y\pi (\phi_{+}^{(0)}-\phi_-)\right)\cos\left(y\pi(\phi_{+}-\phi_-)\right) e^{-\frac{y^2 D^2}{2}N}dy \, , \quad     (b\rightarrow\infty) 
   \end{split}
\end{equation}
 in which  we have approximated the upper bound of the integral by infinity as the exponential function decays rapidly for $|y|>1$. Taking the integral on the right hand side of Eq. \eqref{fpint} we then obtain,
\begin{equation}
    b_{+}(\phi_+,N)=\frac{1}{\sqrt{2 \pi D^2 N }}\Big[e^{-\frac{(\phi_{+}^{(0)}+\phi_+-2 \phi_-)^2}{2 D^2 N}}+e^{-\frac{(\phi_{+}^{(0)}-\phi_+)^2}{2 D^2 N}}\Big] \, ,
  \quad  \quad     (b\rightarrow\infty)  \, .
\end{equation}

%%%%%%%%%%%%%%%%%%%%%%%%%%%%%%%%%%%%%%%
\begin{figure} \label{PDF-free}
    \centering
    \includegraphics[scale=0.45]{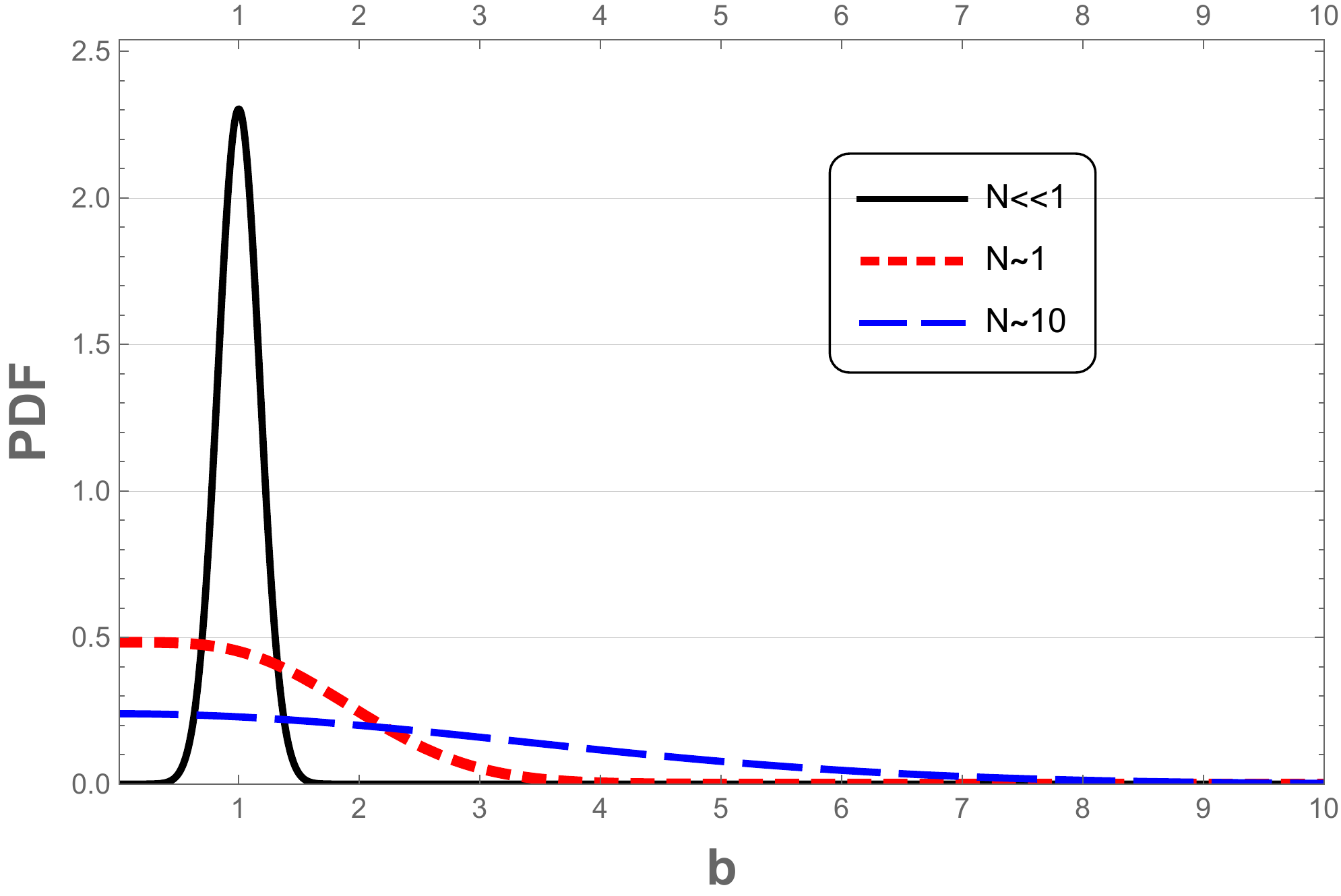}
    \caption{The probability density function for a Brownian boundary without a right barrier ($b \rightarrow \infty$).  This figure is plotted for $\phi_-=0$, $\phi_{+}^{(0)}=1$ and $D=1$. It can be seen that at very early time, $N\ll1$, the field has a Gaussian distribution with a maximum around $\phi_{+}^{(0)}$. However, as time passes, the maximum of distribution is shifted towards $\phi_-$ which is a consequence of $\phi_-$ being a reflective boundary. }
  \label{PDF-free}
\end{figure}
%%%%%%%%%%%%%%%%%%%%%%%%%%%%%%%%%%%%%%%

The behaviour of the above PDF as a function of $\phi_+$ is presented in Fig. \ref{PDF-free}.
This figure is plotted for $\phi_-=0$ and $\phi_{+}^{(0)}=1$. It can be seen that at very early time, $N\ll1$, the field has a Gaussian distribution with a maximum around $\phi_{+}^{(0)}$. This is what one expects since enough time has not passed and therefore the boundary $\phi_+^{(0)}$ is still near its initial value. However, as time goes by, the maximum of distribution is not located around $\phi_+^{(0)}$ anymore and it is shifted towards $\phi_-$ which is a consequence of $\phi_-$ being a reflective boundary. This also confirms that the stochastic boundary forgets its initial value after a while and its distribution is almost uniformly spread along the whole range of $b$.
Also, one can easily check that the above PDF is normalized in the range $(\phi_-,\infty)$ for $N>0$.

Now we obtain the PDF of the field $\phi$.  Suppose that at the initial time $N_0$ the field is located at $\phi_0$. Denoting the corresponding PDF of the Brownian motion by $f(\phi,N|\phi_0,N_0)$, the Fokker-Planck equation governing the stochastic dynamics of the field is given by 
\begin{equation}
   \frac{\partial f(\phi,N|\phi_0,N_0)}{\partial N}=\frac{1}{2}\frac{\partial^2 f(\phi,N|\phi_0,N_0)}{\partial \phi^2},
\end{equation}
whose solution (subject to the initial condition)  is given by
\begin{equation}\label{f-field}
f(\phi,N|\phi_0,N_0)=\frac{1}{\sqrt{2\pi(N-N_0)}}\exp\Big(-\frac{(\phi-\phi_0)^2}{2(N-N_0)}\Big).
\end{equation}

Having at hand the PDF of the boundary we can obtain an equation for the conditional probabilities to hit each of the boundaries at a fixed time $N$. Let $\gamma^+ (N|\phi_0,N_0 )$ $\big(\gamma^-(N|\phi_0,N_0) \big)$ denotes the first  time distribution to hit $\phi_{+}(N)$($\phi_-(N)$) by the condition that   $\phi_{+}(N)$ ($\phi_-(N)$) is crossed earlier than $\phi_-(N)$ ($\phi_{+}(N)$). Note that these two functions are not normalized to unity and they satisfy,
 \begin{equation}
    \int dN \gamma^{\pm}(N|\phi_0,t_0)=p_\pm ,
 \end{equation}
in which $p_+(p_-)$ is the first hitting probability to the right (left) boundary. 
 
 Using the method of Volterra integral equations as in  \cite{Bou:1990} one can show that $\gamma^\pm(N|\phi_0,N_0)$ satisfy the following integral relations (see Appendix \ref{proof} for further details),
\begin{equation}\label{gammaminus}
\begin{split}
    \gamma^-(N|\phi_0,N_0)&=2\psi(\phi_-,N|\phi_0,N_0)\\&\, - 2\int^N_{N_0}dt\Big(\gamma^-(t|\phi_0,N_0)\psi(\phi_-,N|\phi_-,t)+\gamma^+(t|\phi_0,N_0)\psi(\phi_-,N|\phi_+,t)\Big),
 \end{split}
\end{equation}
\begin{equation}\label{gammaplus}
\begin{split}
    \gamma^+(N|\phi_0,N_0)&=-2\psi(\phi_+,N|\phi_0,N_0)\\&+ 2\int^N_{N_0}dt\Big(\gamma^-(t|\phi_0,N_0)\psi(\phi_+,N|\phi_-,t)+\gamma^+(t|\phi_0,N_0)\psi(\phi_+,N|\phi_+,t)\Big),
    \end{split}
\end{equation}
where $\psi(x,N|y,t)$ is defined as 
\begin{equation}
    \psi(x,N|y,t)\equiv\frac{d}{dN}F(x,N|y,t),
\end{equation}
and $F(x,N|y,t)$ is the transition function of $\phi$ defined by
\begin{equation}
    F(x,N|y,t)\equiv P(\phi\leq x,N|y,t)=\int^x_{-\infty}f(\phi,N|y,t)d\phi.
\end{equation}
The proofs of Eqs.(\ref{gammaplus}) and (\ref{gammaminus}) are presented in more details   in the Appendix \ref{proof}.
%%%%%%%%%%%%%%%%%%%%%%%%%%%%%%%%%%%%%%%%%%%%%%%%%%%%%%%%%%%%%%%%%%%%%%%%%%%%%%%%%% 
\subsection{Boundary Crossing Probabilities}
\label{Boundary-Crossing}

In this subsection we find a solution for $p_\pm$, i.e the probabilities to cross a boundary earlier than the other one. Now recall from the above discussions 
that $p_\pm$ are given by, 
\begin{equation}
    p_\pm=\int_0^\infty\gamma^\pm(t)dt \, .
\end{equation}
As it can be seen from Eqs \eqref{gammaminus} and \eqref{gammaplus} the full analytic solution of $\gamma^\pm(t)$ is not possible so we look for approximate ones. For this purpose, we take the Laplace transformations of Eqs. \eqref{gammaminus} and \eqref{gammaplus}. If $\Gamma^\pm(s)$ denote the Laplace transformation of $\gamma^\pm(t)$ then
\begin{equation}
    \Gamma^\pm(s)=\int_0^\infty e^{-st}\gamma^\pm(t)dt.
\end{equation}
It is easy to see that $p_\pm=\lim_{s\rightarrow0}\Gamma^\pm(s)$.  

In the Appendix \ref{prob} we have presented the equations to solve the Laplace transformation of $\gamma^\pm(t)$. Using Eqs. (\ref{gamapluseq}) and (\ref{gamaminuseq}) for $\Gamma^\pm(s)$ and taking the limit $s\rightarrow0$ we obtain
\begin{equation}
\label{pplus}
\begin{split}
    p_+&=1-p_-\\&=\frac{2 \left(\phi _0-\phi_-\right)}{b-2 \phi_-}-4b\sum_{m=1}^\infty\Bigg[\frac{ (-1)^m-1 }{m^2\pi ^2  (b-2\phi_-)}\Bigg] 
\Gamma^+\Big(\frac{ m^2\pi ^2 D^2}{2b^2}\Big)\cos \left(\frac{m\pi }{b}\phi_{+}^{(0)}\right).
\end{split}
\end{equation}
This provides a ``formal'' solution for $p_\pm$. However, to find their actual values we still need to calculate the unknown function $\Gamma^+(\frac{ m^2\pi ^2 D^2}{2b^2})$. As we shall see below, this can be done only iteratively by setting 
$m=1, 2,...$ and then finding the values of $p_\pm$ at the corresponding order
of $m$.  The first term in Eq. (\ref{pplus}) represents the contribution from the leading order term of the PDF ($m=0$). In this case, one can imagine that the right boundary  on average is fixed in the midpoint $\phi_+=\frac{b}{2}$ as it has equal chances to be either to the right or to the left of the point  $\phi_+=\frac{b}{2}$. 

Before solving for $p_\pm$ iteratively, we note that due to the exponential suppression of PDF in Eq. \eqref{fp}, one expects that the solutions for 
$p_\pm$ converge rapidly for large enough $m$ or small value of $\frac{b}{D}$.  We can estimate the rate of convergence by noting that if $\frac{ m^2\pi^2 D^2}{2b^2} \gg 1$ then the exponential terms are negligible. This condition is equivalent to
\begin{equation}
\label{m-b-condition}
    m \gg \frac{b}{\pi D} \, .
\end{equation}
So we see that if $\frac{b}{D} \ll 1$ then after few terms the series is near its final value while for larger values of $\frac{b}{D}$ we should take into account more terms in the series expansion.

To set our convention, we denote the leading order expansion corresponding to case $m=0$ by LO. The next leading expansion, corresponding to $m=1$, is denoted by NLO while the cases of higher orders in $m$ are denoted by $N^mLO$. For example, the next to NLO order with $m=2$ is represented by $N^2LO$. 

At LO order, the solution for $p_+$ from \eqref{pplus} is given by 
\begin{equation}
\label{pLO}
    p_+^{LO}=\frac{2 \left(\phi _0-\phi_-\right)}{b-2 \phi_-} \, .
\end{equation}
This expression is consistent with the result in the case of two fixed boundaries located at $\phi_-$ and $\phi_+ = \frac{b}{2}$ \cite{Firouzjahi:2020jrj}. As discussed above, this makes sense since for the LO order   the right boundary  on average is  fixed in the midpoint $\phi_+=\frac{b}{2}$.

Now, we proceed to calculate $p_+$ at NLO order. To this end one has to calculate $\Gamma^+(\frac{\pi^2 D^2}{2b^2})$. We should also note that Eq. \eqref{gamapluseq} holds for any $s\geq0$. By evaluating $\Gamma^+(\frac{\pi^2 D^2}{2b^2})$ from this equation, one can calculate $p_+$ up to NLO order as follows,
\begin{equation}\label{PPNLO}
    p_+^{NLO}= p_+^{LO}+\frac{8 \,b}{\pi ^2 \left(b-2 \phi _-\right)}\cos \big(\frac{\pi  \phi_{+}^{(0)}}{b}\big)\,Y  ,
\end{equation}
in which
\begin{equation}\label{Y}
\begin{split}
    Y& \equiv
    \Bigg(\frac{\pi D}{\pi ^2 D^2-4  e^{\pi D \left(\frac{2 \phi _-}{b}-1\right)}\sinh ^2\left(\frac{\pi D }{2}\right)}\Bigg) \Bigg[\left(e^{\pi D }-1\right)   e^{-\frac{\pi D}{b}  \left(b-2 \phi _-+\phi _0\right)} \left(e^{\frac{2 \pi D}{b}  \left(\phi _0-\phi _-\right)}-1\right)\\&-2 D^2 \sum_{j=0}^{\infty}\Big[\frac{(-1)^{j} - e^{\pi D  \sqrt{ j^2+1}}}{\left((1+D^2)j^2+D^2\right)  \sqrt{j^2+1} }\Big]\cos\left(\frac{j\pi }{b}{\phi_{+}^{(0)}}\right) e^{\frac{\pi D}{b} \sqrt{j^2+1} \left(\phi _0-b\right)}\Bigg] \, .
\end{split}
\end{equation}
We can go one step further and calculate $p_+$ at the $N^2LO$ order, corresponding to $m=2$. However, the results for $N^2LO$ order are very complicated and we avoid presenting them here. 

\begin{figure}
    \centering
    \begin{subfigure}[t]{0.48\textwidth}
        \centering
        \includegraphics[width=\linewidth]{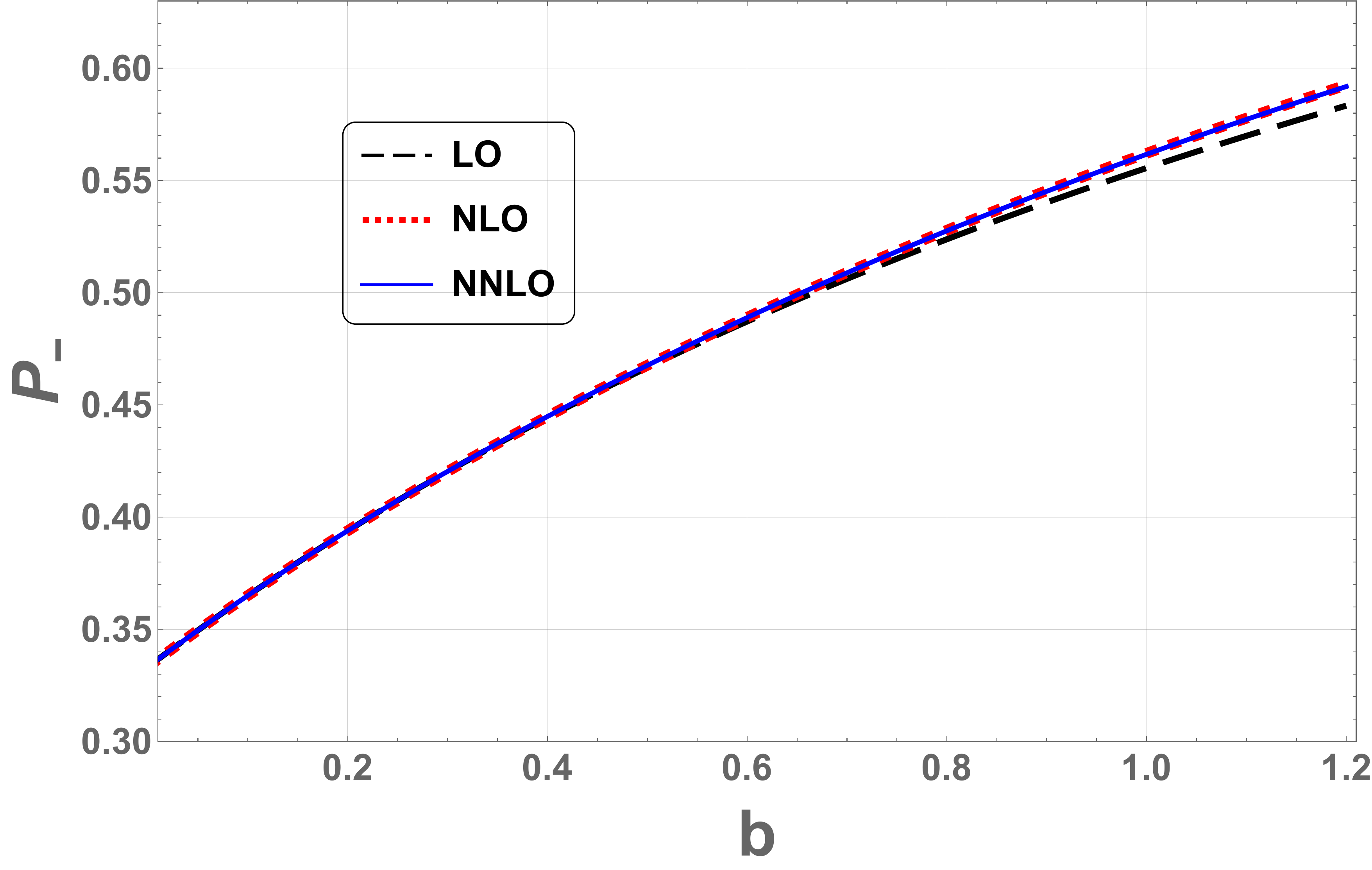} 
        \caption{} \label{Pm-b}
    \end{subfigure}
    \hfill
    \begin{subfigure}[t]{0.48\textwidth}
        \centering
        \includegraphics[width=\linewidth]{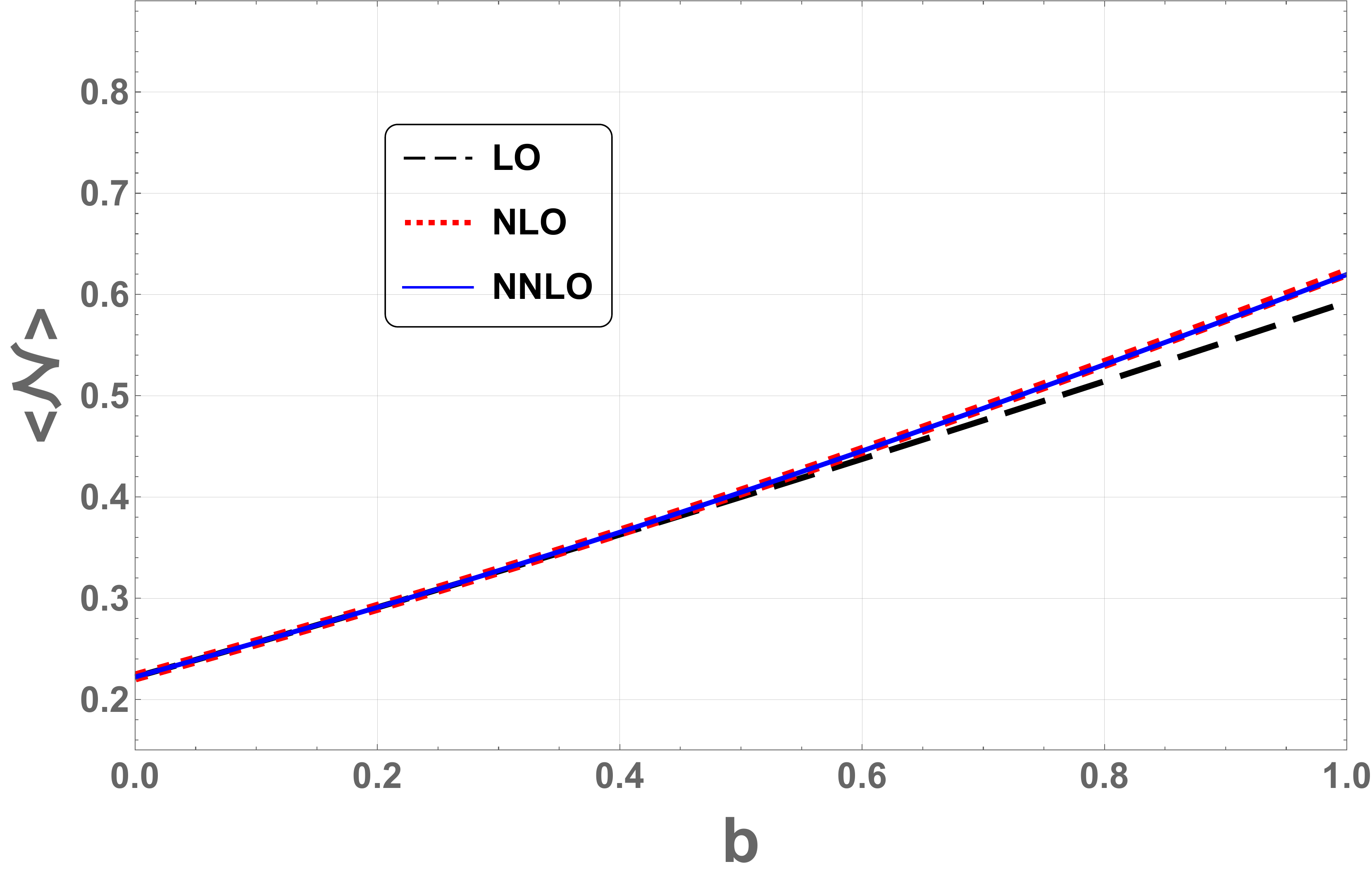} 
        \caption{} \label{N-b}
    \end{subfigure}
    \caption{The behaviour of $p_-$ (left)  and ${\langle{\cal{N}\rangle}}$ (right) versus $b$ for LO, NLO and $N^2LO$ terms.  Here we have set $\phi_-=-1$, $\phi_0=\frac{-1}{3}$,  $\phi_{+}^{(0)}=0.7b$ and $D=1$. As one expects, as $b$  increases  $p_+$ is decreased and correspondingly  $p_-$ is increased. Furthermore, as 
  $b$ increases the average time $\langle \cN \rangle$ for the field to hit either of the boundaries increases as well.}
    \label{Pm-x03}
\end{figure}

\begin{figure}
 \vspace{1cm}
    \centering
    \begin{subfigure}[t]{0.48\textwidth}
        \centering
        \includegraphics[width=\linewidth]{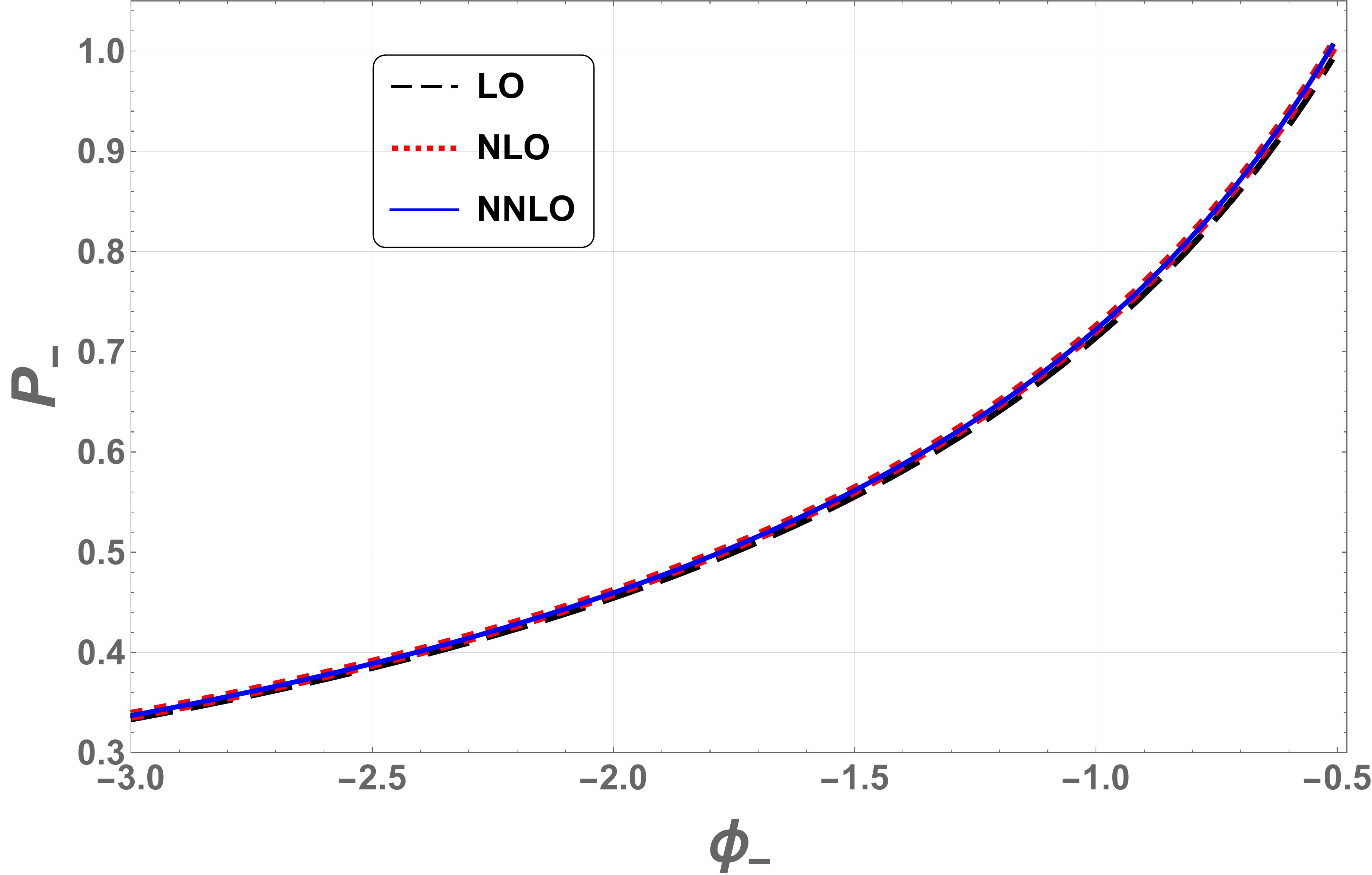} 
        \caption{} 
    \end{subfigure}
    \hfill
    \begin{subfigure}[t]{0.48\textwidth}
        \centering
        \includegraphics[width=\linewidth]{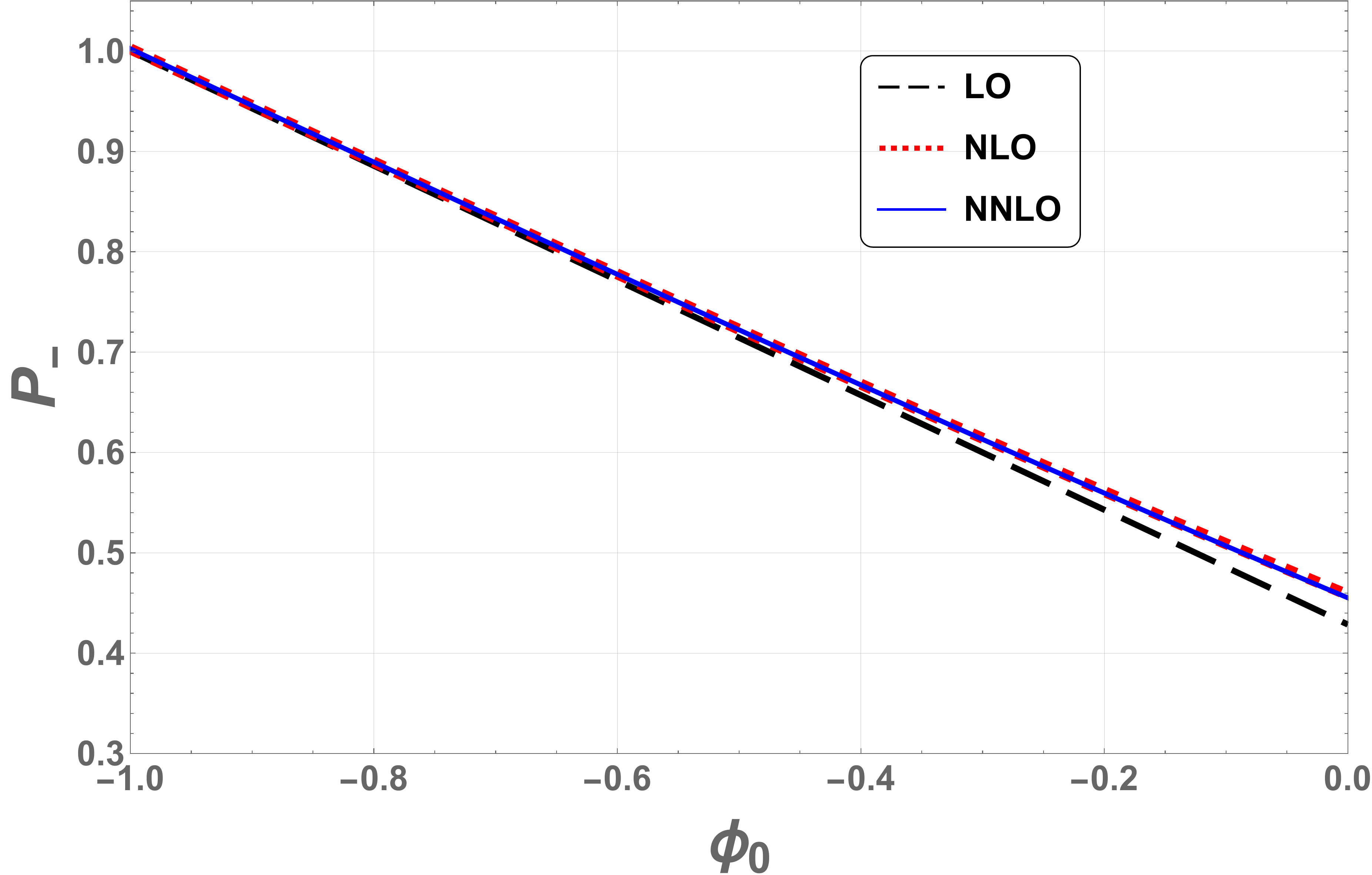} 
        \caption{} 
    \end{subfigure}
    \caption{A comparison of $p_-$ versus $\phi_-$ (left) and $\phi_0$ (right) for LO, NLO and $N^2LO$. Here we have set $b=1.5$, $\phi_{+}^{(0)}=0.7b$ and $D=1$. In the left panel we have considered $\phi_0=-\frac{1}{2}$ and as one expects, when $\phi_-=\phi_0=-\frac{1}{2}$ i.e. when  the field is located on the position of left boundary, the probability equals to unity. However, as
 $\phi_-$ moves away from $\phi_0$ then $p_-$ is reduced. In the right panel, however, we have set $\phi_- = -1$ in which for $\phi_0 = -1$ one obtains $p_- = 1$ as expected. As $\phi_0$ moves away from this value the hitting probability $p_-$ is decreased. }
    \label{Pm-x0}
\end{figure}

As the condition (\ref{m-b-condition}) indicates  the above result for $p_+^{NLO}$ 
is a good approximation to the value of $p_+$  as far as $\frac{b}{D}\lesssim 1$ while for large values of $\frac{b}{D}$ there can be significant deviations.  As $\frac{b}{D}$ gets larger and larger, the higher order terms in the series with $m \ge 2$ become  non-negligible. In Figs.  \ref{Pm-x03} and \ref{Pm-x0}  we have compared the LO, NLO and $N^2LO$ terms for $p_-$ for different variables. Fig. \ref{Pm-b} shows the behaviour of $p_-$ with respect to $b$ which is plotted at various orders  for $\phi_-=-1$ and $\phi_0=\frac{-1}{3}$. In addition, we have set the initial value of the stochastic boundary to be $\phi_{+}^{(0)}=0.7b$ so it is a function of $b$ as $b$ varies. 
 As one expects and can be seen from this figure, increasing $b$ (with a fixed value of $D$) results in a higher value for $p_-$, i.e higher probability of hitting the left boundary corresponding to a lower probability of hitting the right boundary. In addition, the convergence in the series expansion 
 is fast as the curves representing the plots of $p_-$ for 
 NLO and $N^2LO$ orders are nearly identical. 
 
In Fig. \ref{Pm-x0} the behaviours of $p_-$ with respect to $\phi_-$ and $\phi_0$ are plotted at various orders. One can see from the left panel of this figure that 
 with $b$ and $D$ kept fixed, as $\phi_-$ moves  away from $\phi_0$, the probability of hitting the left boundary ($p_{-}$) decreases which is expected. Also, the right panel shows the behaviour of $p_-$
versus the initial values of the field $\phi_0$ for fixed values of 
$b, D$, $\phi_{+}^{(0)}$ and $\phi_{-}$.  As can be seen from this panel, for $\phi_0=\phi_-=-1$, i.e. when the field is initially located on the location of boundary,  the probability of hitting $\phi_-$ is equal to unity and as the initial position of the field moves away from $\phi_-$ this probability decreases.

We comment that  the result \eqref{pplus} can be obtained via another independent method as we elaborate below.  Taking the average of Eq. \eqref{langevinphi},  yields 
\begin{equation}\label{AL}
    \left\langle \phi-\phi_0\right \rangle=p_+ E(\phi-\phi_0|+)+p_-E(\phi-\phi_0|-)=0,
\end{equation}
where $E(\phi|+)$ \big($E(\phi|-) \big)$ represents the conditional average value of the field by the condition that $\phi_{+}(N)$\big($\phi_-(N)$\big) is crossed earlier than $\phi_-(N)$\big($\phi_{+}(N)$\big). The following expressions hold:
\ba
\label{ppEplus1}
  p_+ E(\phi-\phi_0|+) &=&\int_{0}^{\infty} \int_{0}^{b} d\tau dx \gamma^+(\tau|\phi_0,t_0)(x-\phi_0) f_{+}(x,\tau),\\
%\end{equation}
%\begin{equation}
\label{pminusE}
    p_- E(\phi-\phi_0|-)&=&p_-(\phi_--\phi_0) \, .
\ea
Now note that Eq. \eqref{ppEplus1} can be written as 
%\begin{equation}
\ba
\label{pplusE}
%\begin{split}
    p_+ E(\phi-\phi_0|+) &=& 
    \int_0^\infty d\tau \, \gamma^+(\tau|\phi_0,t_0)  \nonumber\\
   &\times&   \int_{0}^{b} dx\, (x-\phi_0) \left[\frac{1}{b}+\frac{2}{b}\sum_{m=1}^{\infty}\cos\left(\frac{m\pi}{b} \phi_{+}^{(0)}\right)\cos\left(\frac{m\pi }{b}x\right) e^{-\left(\frac{ m^2\pi^2 D^2 }{2 b^2}\right)\tau}\right] \, , \nonumber\\
    &=& p_+\Big(\frac{b-2\phi_0}{2}\Big)+2b\sum_{m=1}^\infty \left[\frac{(-1)^m-1}{m^2\pi^2}\right]\cos\Big(\frac{m\pi}{b^2} \phi_{+}^{(0)}\Big)\Gamma^+\Big(\frac{ m^2 \pi^2 D^2}{2b^2}\Big) \, .
    %\end{split}
%\end{equation}
\ea
Substituting Eqs. \eqref{pminusE} and \eqref{pplusE} into Eq. \eqref{AL} one obtains the same expression for $p_+$ as Eq. \eqref{pplus}. 

%%%%%%%%%%%%%%%%%%%%%%%%%%%%%%%%%%%%%%
%%%%%%%%%%%%%%%%%%%%%%%%%%%%%%%%%%%%%%

%%%%%%%%%%%%%%%%%%%%%%%%%%%%%%%%%%%%%%

%%%%%%%%%%%%%%%%%%%%%%%%%%%%%%%%%%%%%%

%%%%%%%%%%%%%%%%%%%%%%%%%%%%%%%%%%%%%%
\subsection{Mean Number of e-folds}

Using the same approach we can calculate the mean number of e-folds $\big\langle \mathcal{N}\big\rangle $ for the field to cross either of the boundaries. As mentioned in previous section, while the clock $N$ is a deterministic variable but $\cN$, the number of e-folds hitting either of the boundaries, is a stochastic variable.  
By taking the average of the square of the Langevin equation one obtains,
\begin{equation}
    \big\langle (\phi-\phi_0)^2\big\rangle=p_+E\big((\phi-\phi_0)^2|+\big)+p_-E\big((\phi-\phi_0)^2|-\big)=\big\langle W(\mathcal{N})^2\big\rangle=\big\langle\mathcal{N}\big\rangle \, .
\end{equation}
Now similar to  Eq. \eqref{ppEplus1} and using the results obtained in the Appendix \ref{prob} one can write
\begin{equation}
\label{ppEplus}
\begin{split}
    p_+& E\big((\phi-\phi_0)^2|+\big)=\int_{0}^{\infty} \int_{0}^{b} d\tau dx \gamma^+(\tau|\phi_0,t_0)(x-\phi_0)^2 f_{+}(x,\tau) \, ,\\&=p_+\Big(\frac{1}{3}b^2+\phi _0 \big(\phi _0-b\big)\Big)+\sum_{m=1}^\infty\frac{4 b}{m^2\pi^2}\Big[{(b-\phi _0) \cos (\pi  m)+\phi _0} \Big]\cos\Big(\frac{m \phi_{+}^{(0)}}{b}\Big)\Gamma^+\Big(\frac{ m^2\pi^2 D^2}{2b^2}\Big),
    \end{split}
\end{equation}
and
\begin{equation}
\label{ppEminus}
  p_- E\big((\phi-\phi_0)^2|- \big)=p_-(\phi_{-}-\phi_0)^2. 
\end{equation}

Having the solutions of $p_\pm$ from the previous analysis, we can calculate 
$\langle \cN \rangle $ from Eqs. (\ref{ppEplus}) and  (\ref{ppEminus}). However, as in the case of $p_\pm$, 
the result for $\langle \cN \rangle $ can be obtained only iteratively in series expansion. More specifically, using the LO  expressions for $p_+$ in Eq. (\ref{pLO}), 
 $\big\langle \mathcal{N}\big\rangle$ at LO order is obtained to be, 
\begin{equation}
    {\langle{\cal{N}\rangle}^{LO}}=\frac{\left(\phi _0-\phi _-\right)}{3 \left(b-2 \phi _-\right)} \left[2 b^2-3 b \left(\phi _-+\phi _0\right)+6 \phi _- \phi _0\right].
\end{equation}
Similarly, proceeding to  $\big\langle\mathcal{N}\big\rangle^{NLO}$ at NLO order we obtain  
\begin{equation}
\label{N-NLO}
\begin{split}
    {\langle{\cal{N}\rangle}^{NLO}}={\langle{\cal{N}\rangle}^{LO}}-\frac{4bY }{3 \pi ^2 \left(b-2 \phi _-\right)}\big[b^2-6 b \phi _-+6 \phi _-^2\big]
   \cos\Big(\frac{\pi\phi_+^{(0)}}{b}\Big),
\end{split}
\end{equation}
in which $Y$ is defined as in Eq. (\ref{Y}). One can also obtain  $\langle \cal{N}\rangle$ at next orders.  However, since it takes a very complicated form, we avoid to present it here.

%%%%%%%%%%%%%%%%%%%%%%%%%%%%%%%%%%%%%%%%
\begin{figure}
    \centering
    \begin{subfigure}[t]{0.48\textwidth}
        \centering
        \includegraphics[width=\linewidth]{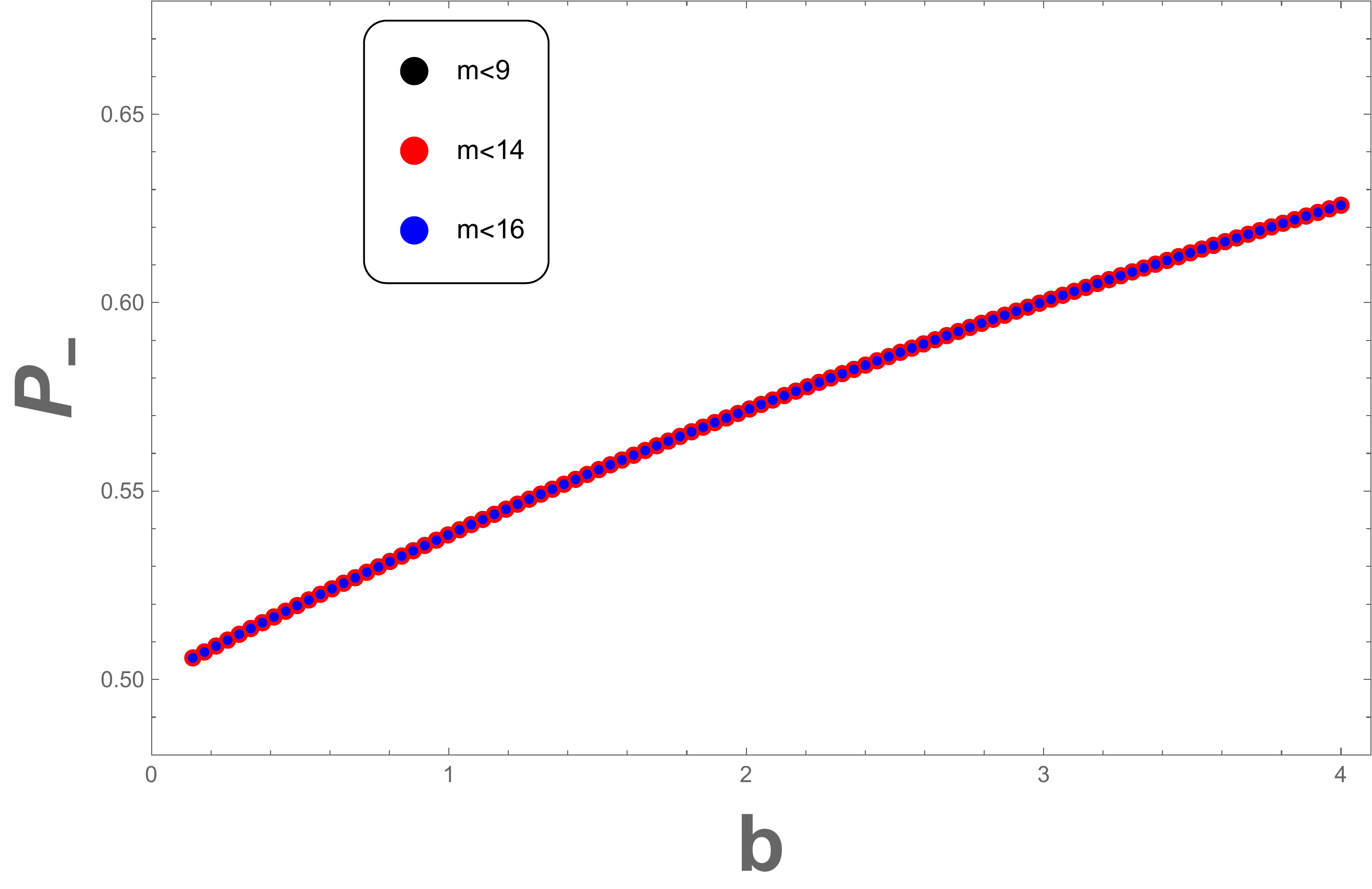} 
        \caption{} \label{Pm-b-M=16}
    \end{subfigure}
    \hfill
    \begin{subfigure}[t]{0.47\textwidth}
        \centering
        \includegraphics[width=\linewidth]{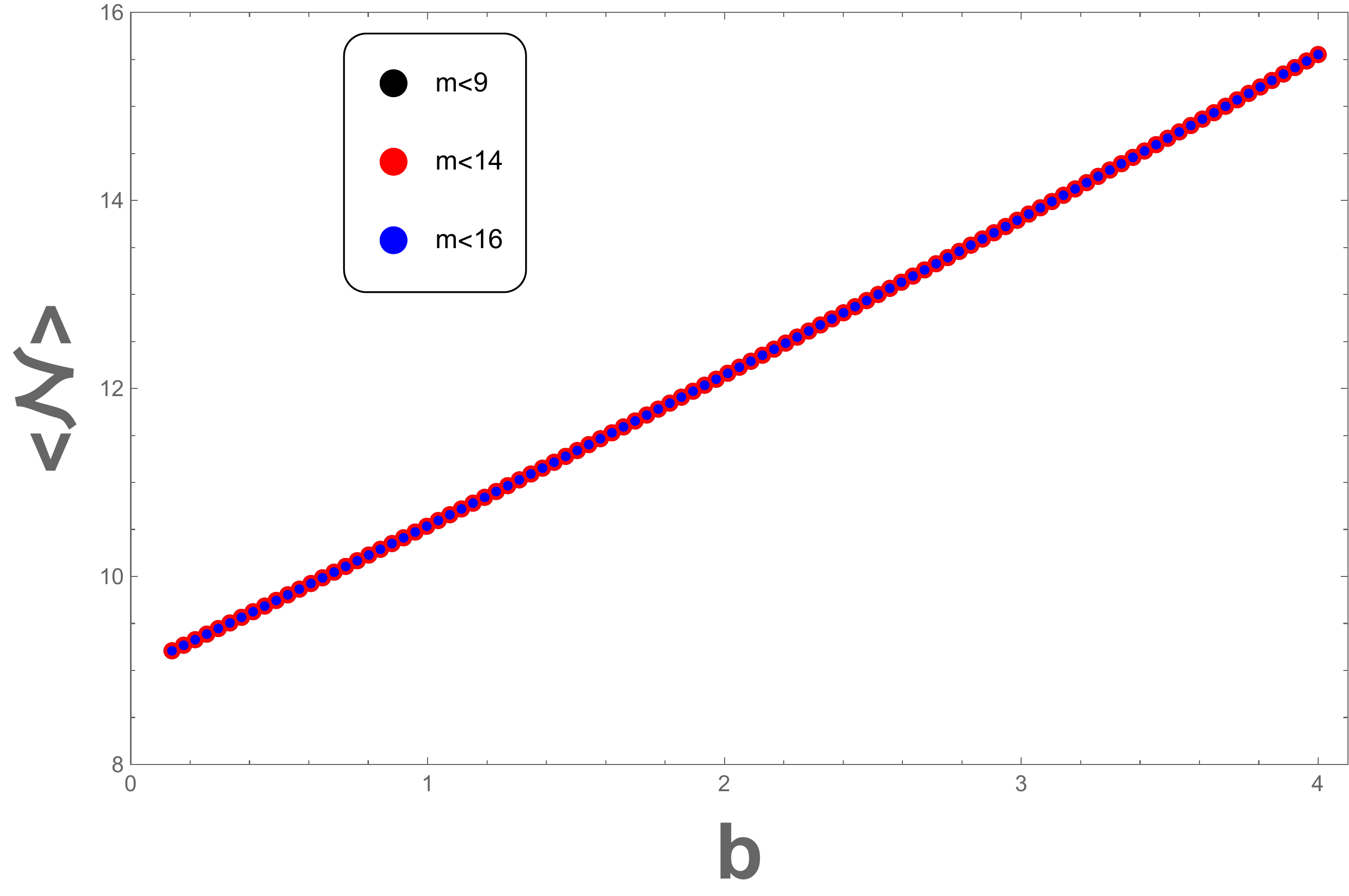} 
        \caption{} \label{N-b-M=16}
    \end{subfigure}
    \caption{The behaviour of $p_-$ (left) and   and $\big<\mathcal{N}\big>$ (right) including higher orders of $m$, for $m<9$, $m<14$ and $m<16$. The initial conditions are $\phi_0=-3$, $\phi_-=-6$, $p_{0}=0.7 b$ with $D=1$. 
 As one expects by increasing $b$ the probability $p_+$ of hitting the stochastic boundary (i.e. right boundary) decreases and consequently $p_-$ increases. Furthermore, the average time it takes  for the field to hit the boundaries increases.}
 \label{various-m}
 \vspace{1cm}
\end{figure}
%%%%%%%%%%%%%%%%%%%%%%%%%%%%%%%%%%%%%%%%

The behaviour  of $\langle \cN \rangle $ with respect to $b$ for  LO, NLO and $N^2LO$ are plotted in  Fig. \ref{N-b}. As  can be seen, by increasing $b$ the average time  the field needs to hit either of the boundaries increases which is what one expects as well. Moreover, the figure shows that for smaller values of $b$, the results for LO, NLO and $N^2LO$  converge very well while for larger values of $b$ the deviations between the  previous orders become more enhanced. This is in line with our conclusion, such as in Eq. (\ref{m-b-condition}),  that 
for large values of $b$ higher orders (i.e. higher orders of $m$) 
should be included in the series expansion for better convergence. 

In Figs.  \ref{Pm-b-M=16} and \ref{N-b-M=16} the plots of $p_-$ and $\langle \cN \rangle$ for higher orders of the series expansion, $m< 9, m< 14$ and $m<16$, are presented. As we see from these plots, the curves for these orders are almost identical to each other. This indicates that the series expansion converges rapidly to its final value after a few orders in expansion. Furthermore, as we observed previously,  the probability of hitting the left 
boundary $p_-$ grows as $b$ increases, confirming the conclusion that the probability of hitting the right boundary $p_+$ is reduced.  Furthermore, Fig. \ref{N-b-M=16} shows that as $b$ increases $\langle \cN \rangle$ increases as well, in agreement with the  conclusion from  Fig. \ref{N-b}.

%%%%%%%%%%%%%%%%%%%%%%%%%%%%%%%%%%%%%%%%
\begin{figure}
 \vspace{1cm}
    \centering
    \begin{subfigure}[t]{0.48\textwidth}
        \centering
        \includegraphics[width=\linewidth]{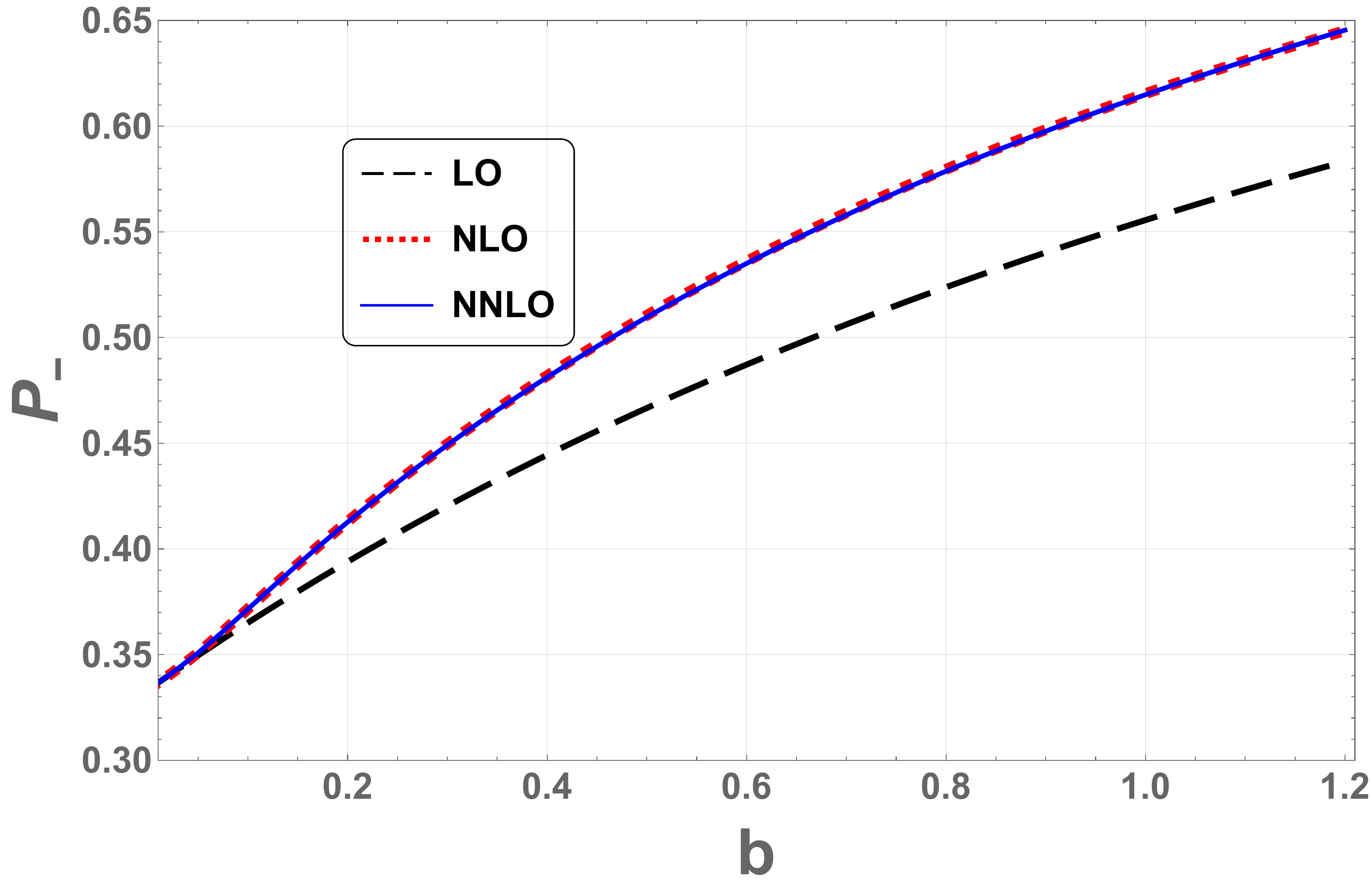} 
        \caption{} 
    \end{subfigure}
    \hfill
    \begin{subfigure}[t]{0.48\textwidth}
        \centering
        \includegraphics[width=\linewidth]{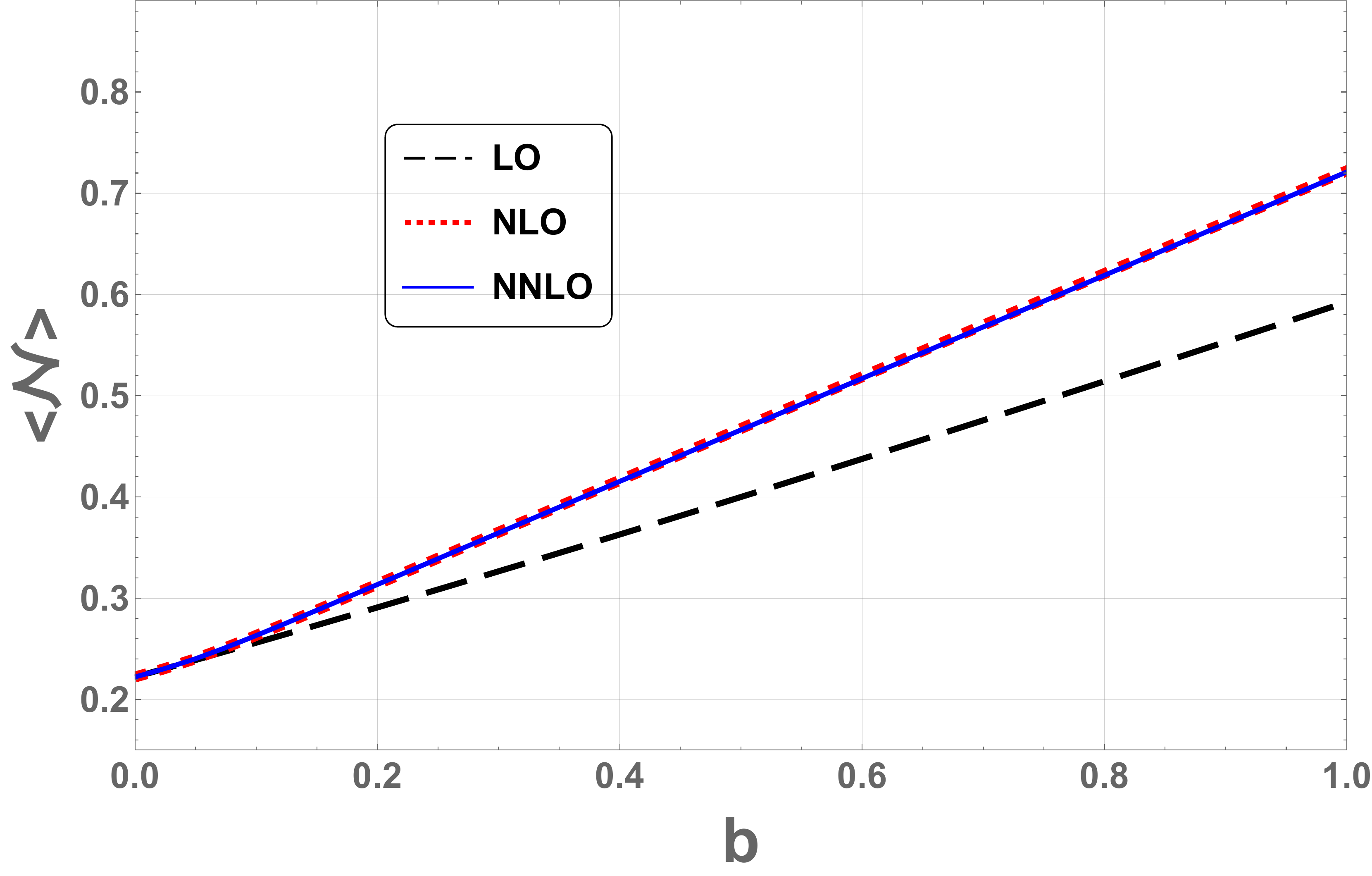} 
        \caption{} 
    \end{subfigure}
    \caption{The behavior of $p_-$ and $\big\langle \mathcal{N}\big\rangle $ versus $b$ for $D=0.1$, $\phi_{+}^{(0)}=0.7b$, $\phi_-=-1$ and $\phi_0=-\frac{1}{3}$.
    This plot is parallel to Fig. \ref{Pm-x03} but now $D$ is reduced by one order of magnitude. As $D$ decreases both $p_-$ and $\big\langle \mathcal{N}\big\rangle $ 
    increase. In addition, the convergence is less efficient as the deviation between 
    LO and NLO orders is intensified compared to the plots in Fig. \ref{Pm-x03}.  }
    \label{pm-SmallD}
    \vspace{1cm}
\end{figure}
%%%%%%%%%%%%%%%%%%%%%%%%%%%%%%%%%%%%%%%%

%%%%%%%%%%%%%%%%%%%%%%%%%%%%%%%%%%%%%%%%

\begin{figure}
    \centering
    \begin{subfigure}[t]{0.48\textwidth}
        \centering
        \includegraphics[width=\linewidth]{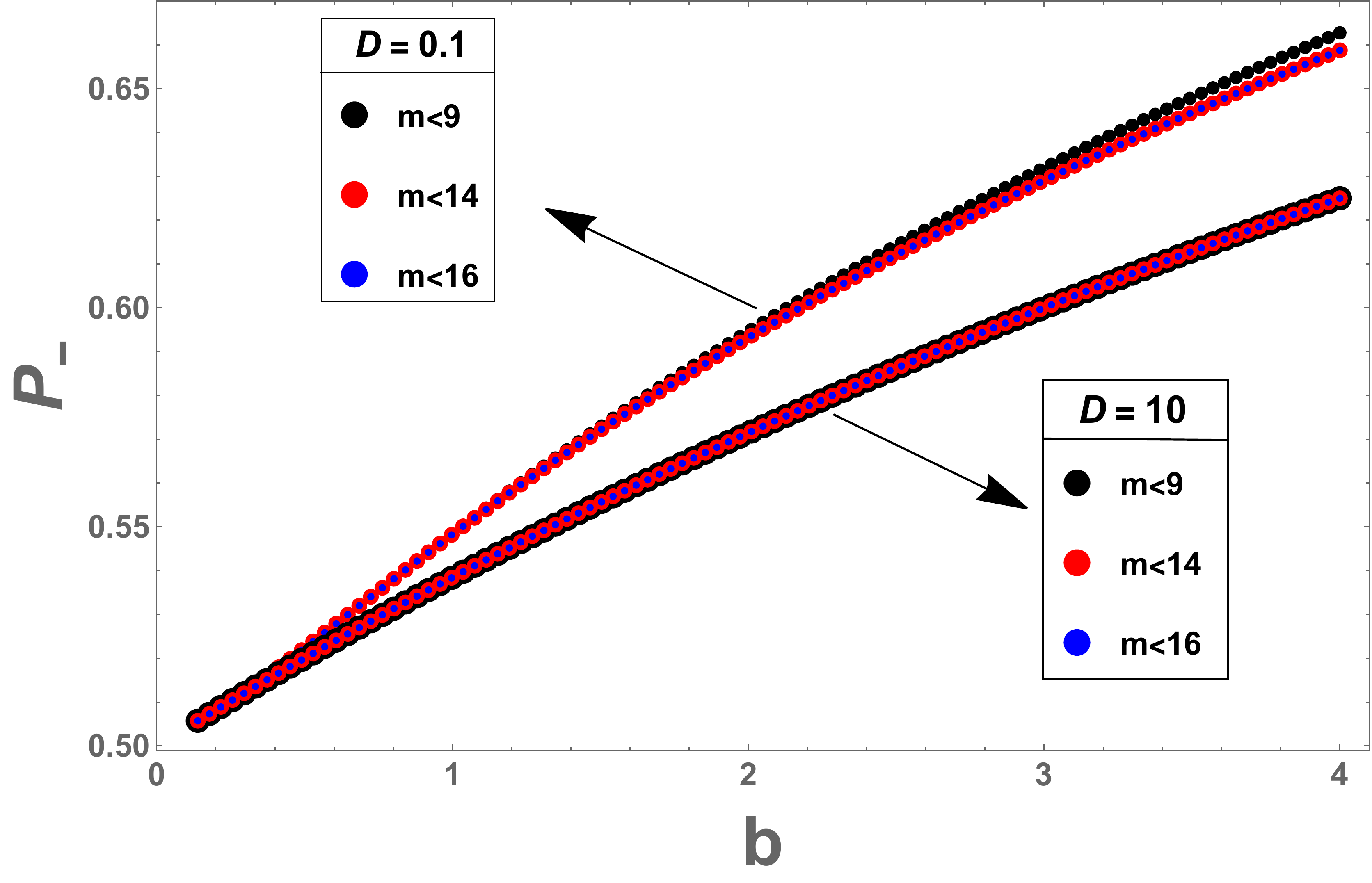} 
        \caption{} 
    \end{subfigure}
    \hfill
    \begin{subfigure}[t]{0.47\textwidth}
        \centering
        \includegraphics[width=\linewidth]{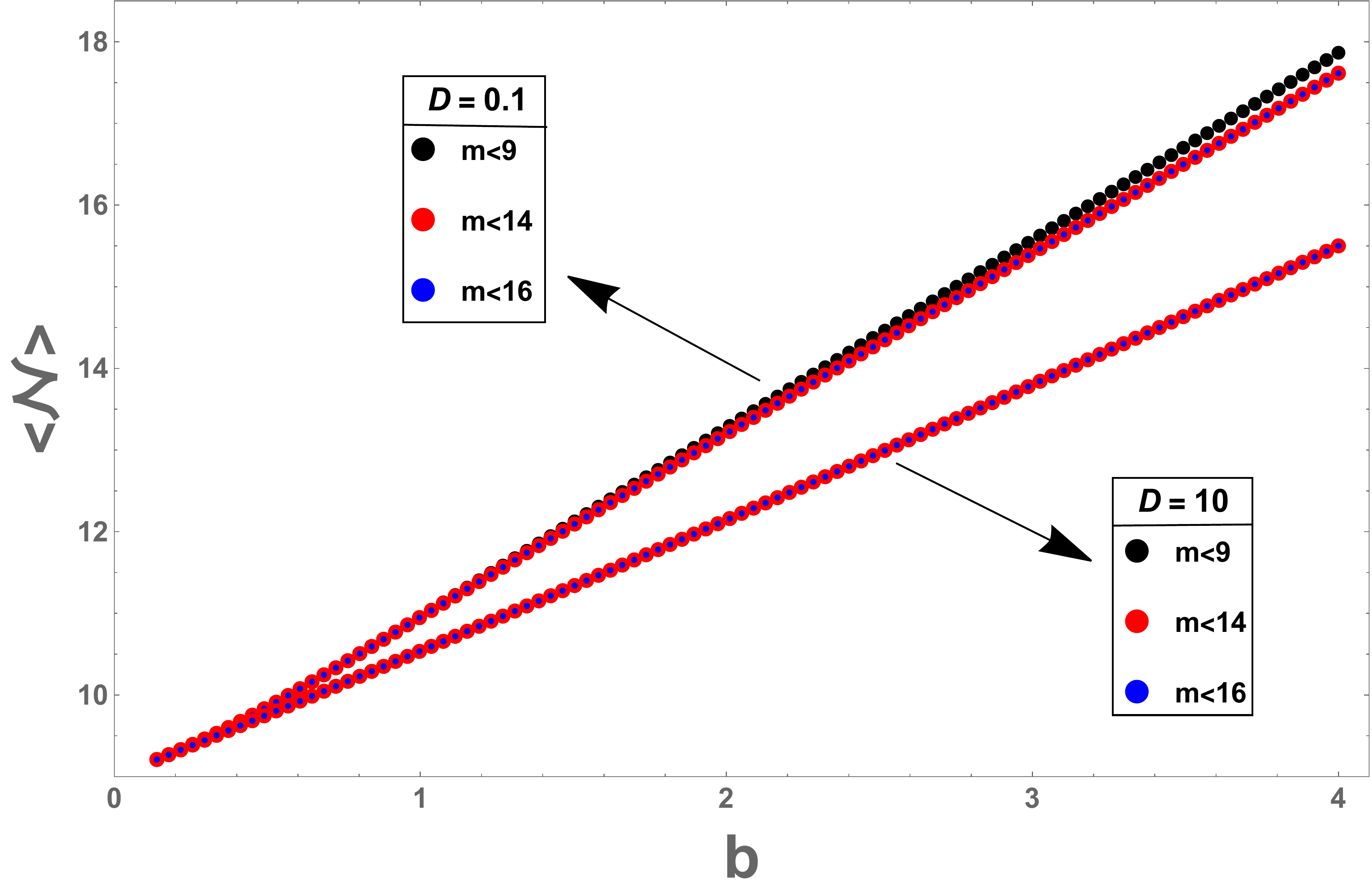} 
        \caption{} 
    \end{subfigure}
    \caption{The behavior of $p_-$ and $\langle \cN \rangle $ versus $b$ for $D=0.1$ and $D=10$. Other initial conditions are the same as in Fig. \ref{various-m}. The conclusion is that as $D$ decreases (increases), both $p_-$ and $\langle \cN \rangle $ increase (decrease). Furthermore, convergence happens efficiently for larger values of $D$. }
    \label{LS}
\end{figure}

In the previous plots, we have set $D=1$, i.e. the field and the stochastic boundary had equal diffusion amplitudes (in the unit of $\frac{H}{2 \pi}$).
However, as we discussed below  Eq. (\ref{m-b-condition}), $D$ like $b$ has important effects on the physical results while the series convergence depends on the combination $\frac{b}{D}$. Intuitively speaking, a larger value of $D$ works parallel to small values of $b$ and vice versa. Specifically, for  $D \ll1$ one needs to add more series terms for the convergence to occur while for $D \gg 1$ the convergence occurs rapidly. These conclusions can be seen in Figs. \ref{pm-SmallD} and   \ref{LS}.
Fig. \ref{pm-SmallD} is a repetition of Fig. \ref{Pm-x03}  but with $D$ decreased 
by one order of magnitude to $D=0.1$. The deviation between the LO and NLO orders in Fig. \ref{pm-SmallD} is intensified compared to Fig. \ref{Pm-x03}.  
On the other hand,  Fig. \ref{LS} is a repetition of  Fig. \ref{various-m}. As we see, 
the cases with small values of $D$ show deviations in series convergence even for large values of $m$ (here the case $m<9$)  while there is no such effect for 
$D \gg 1$. In addition, we note that as $D$ decreases (increase), both $p_-$ and $\langle \cN \rangle$ increase (decrease). This is because a small $D$ represents a small Brownian jump for the boundary so its takes many steps from the field to hit the boundary. Also note that this conclusion is consistent with our intuition that the effects of $D$ work opposite to the roles of $b$. Consequently, a setup with a small value of $D$ is like a setup with a large value of $b$ yielding to larger values of  $p_-$ and $\langle \cN \rangle$.

%%%%%%%%%%%%%%%%%%%%%%%%%%%%%%%%%%%%%%%%

\begin{figure}[t]
    \centering
    \begin{subfigure}[t]{0.49\textwidth}
        \centering
        \includegraphics[width=\linewidth]{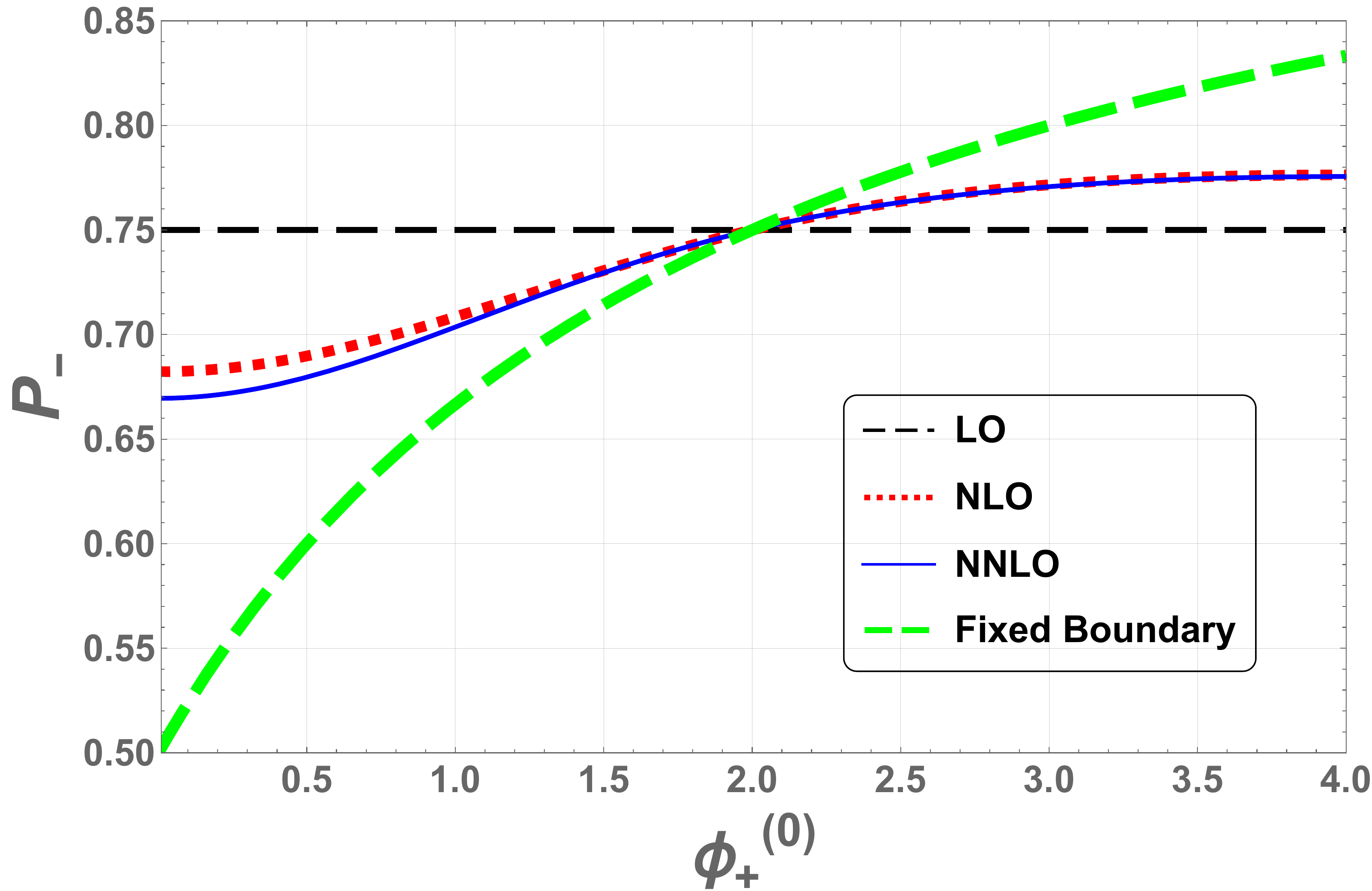} 
        \caption{} \label{Pm-p0}
    \end{subfigure}
    \hfill
    \begin{subfigure}[t]{0.47\textwidth}
        \centering
        \includegraphics[width=\linewidth]{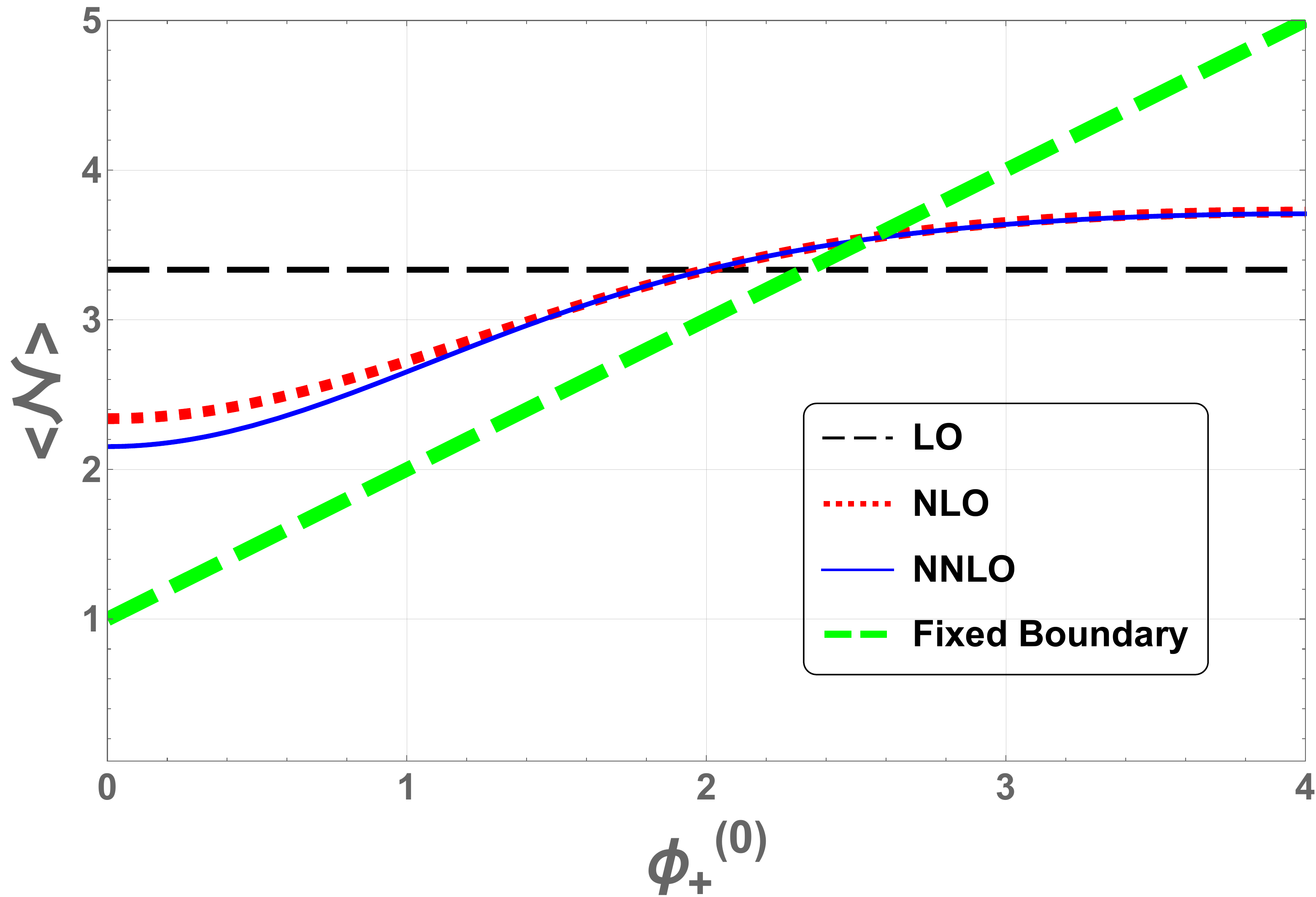} 
        \caption{} \label{N-p0}
    \end{subfigure}
    \caption{Behaviour of $p_-$ (left) and ${\langle{\cal{N}\rangle}}$ (right) versus the initial location of the stochastic boundary, $\phi_{+}^{(0)}$. We have considered $\phi_0=-1$, $\phi_-=-2$, $b=4$ and $D=1$, therefore $\phi_{+}^{(0)}$ can take any value between $0$ and $4$. As we discussed and can be seen from the left panel, for  $\phi_{+}^{(0)}<\frac{b}{2}$, $p_-$ for fixed boundary is less than the stochastic boundary, while for $\phi_{+}^{(0)}>\frac{b}{2}$ it is reversed. 
    }
\end{figure}
%%%%%%%%%%%%%%%%%%%%%%%%%%%%%%%%%%%%%%%%

Finally,  it would be instructive to look at the behaviour of $p_-$ and $\langle \cN \rangle$ as functions of $\phi_{+}^{(0)}$, the initial position of the right boundary.
While keeping all other parameters fixed,  by 
increasing $\phi_{+}^{(0)}$  we expect physically that both $p_-$ and $\langle \cN \rangle$ to increase. This is because the right boundary is moved away further from the field so it takes more time  for the field to hit the right boundary and also with less probability. These conclusions can be seen in  Figs. \ref{Pm-p0} and \ref{N-p0}. 
As a relevant question, it would be interesting to compare the results in the current case where the right boundary itself 
undergoes stochastic motion to the case where the right boundary is fixed, as in the simple example studied in \cite{Firouzjahi:2018vet}. One can show that depending on the initial value of the boundary $\phi_{+}^{(0)}$, the results for $p_-$ and $\langle \cN \rangle$ can be larger or smaller than the scenario with the fixed boundary. To be more precise, first assume $\phi_+^{(0)} < \frac{b}{2}$. In this case, while the right boundary moves stochastically, it can probe the region $\phi_+ > \frac{b}{2}$  as well. So compared to the  case of \cite{Firouzjahi:2018vet} where the boundary is located at a fixed  position $\phi_+< \frac{b}{2}$, the boundary in the current case has more room to go beyond the region $\phi_+ > \frac{b}{2}$. Correspondingly, both $p_-$ and $\langle \cN \rangle$ increase compared to the case of fixed boundary. As an example, like in plots (\ref{Pm-p0}) and (\ref{N-p0}), 
consider the configuration where  the stochastic boundary is initially located at $\phi_+^{(0)}=0.3 b$ and compare it with the case where the boundary is held fixed at $\phi_+=0.3 b$.  Considering both cases, we obtain $\big<\mathcal{N}\big>_{\mathrm{stoc}}=3.02$ and  $\big<\mathcal{N}\big>_{\mathrm{fixed}}=2.2$ which  is consistent with what we concluded, that is $\big<\mathcal{N}\big>_{\mathrm{stoc}}\,>\, \big<\mathcal{N}\big>_{\mathrm{fixed}}$. 
On the other hand, for the case $\phi_+^{(0)} > \frac{b}{2}$ the situation is reversed and both $p_-$ and $\langle \cN \rangle$ decrease compared to the fix boundary case. As an example, suppose  $\phi_{+}^{(0)}=0.7b $ with other initial condition as in previous example.  We obtain  $\langle\mathcal{N}\rangle_{\mathrm{stoc}}=3.53$ while $\langle\mathcal{N}\rangle_{\mathrm{fixed}}=3.8$ which is again consistent with our conclusion.  All these interesting properties can be seen in Figs. \ref{Pm-p0} and \ref{N-p0}. Another interesting point is that, according to Eq. (\ref{symmetry}), considering $\phi_+^{(0)}=\frac{b}{2}$ implies that the average time that the field hits the boundary in the interval $0<\phi_+<\frac{b}{2}$ is equal to the case in  which $\frac{b}{2}<\phi_+<b$. This may seem opposite to  one's expectation that the closer  $\phi_+$ is the sooner the crossing time would be. However, we should note that during the crossing time the boundary is not fixed and it can come from different points to the crossing region. 

Up to this point we have assumed that the right boundary undergoes Brownian motion. As we saw from Eq. (\ref{fp}), the LO term in $f_+(\phi,N)$ is a constant which is very similar to a uniform density. However, one should note that its nature is completely different than a boundary with a uniform density. To be more precise, a boundary with a uniform density evolves discontinuously in time while a  boundary with a Brownian motion has a continuous evolution. To see the differences, in the next section  we study the case where the right boundary undergoes stochastic motion with a  uniform distribution.

\section{Boundary with Uniform Distribution}\label{Uniform2}
%%%%%%%%%%%%%%%%%%%%%%%%%%%%%%%%%%%%%%%%%%%%%%%%%%%%%%%%%%%%%%%%%%%%%%%%%%%%%%%%%%

In the previous section we have studied a scalar field with Brownian motion which was restricted between two boundaries, one held fixed while the other one experiencing a pure Brownian motion. However, it would also be interesting to study the case in which the stochastic boundary (i.e. the right boundary) has a uniform distribution. This distribution is represented by the first term of  $f_+(\phi_+,N)$ 
in Eq. (\ref{fp}).  However, due to non-Markovian evolution of the boundary with the uniform distribution, the results for $p_\pm$ and $\langle\mathcal{N}\rangle$ are totally different from the ``LO'' order obtained in previous section.  As the boundary with uniform distribution has a discontinuous evolution with time we can not use Eqs  \eqref{gammaminus} and \eqref{gammaplus} to obtain $p_\pm$ and $\langle\mathcal{N}\rangle$. 

As the boundary in this case does not have a well defined evolution with time we only present the PDF of the boundary:
\begin{equation}
    f_{+}(x,t)=\frac{1}{b} , \quad\quad 0<x<b \, .
\end{equation}
As it does not have a Markovian evolution, the boundary can take any value in the interval  $0<x<b$ at time $t+dt$ regardless of its initial condition at the time $t$. Moreover, the time evolution of the field is the same as Eq. \eqref{langevinphi} with the initial condition $\phi_0$ 
while the left boundary is kept fixed at $\phi_-$.
To obtain $p_\pm$, first we consider $\phi_+$ to be at a position say $y$. Then the hitting probability of the field 
with the initial value $\phi_0$, which is restricted between two fixed boundaries located at 
$\phi_-$ and $y$, is obtained to be $( \frac{\phi_0-\phi_-}{y-\phi_{-}})$ \cite{Firouzjahi:2018vet}. This in turn   gives the following result for $p_+$,
\begin{equation}\label{P-uniform}
\begin{split}
    p_+=&\int_0^b{p(\phi\,\, \text{hits}\,\, \phi_+\,\, \text{first}|\phi_+=y) \,\frac{dy}{b}}\\
    =&\int_0^b \big( \frac{\phi_0-\phi_-}{y-\phi_-}\big)\,\frac{dy}{b}\\
    =&\big(\frac{\phi_0-\phi_-}{b}\big)\ln{\big(\frac{\phi_--b}{\phi_-}\big)}.
    \end{split}
\end{equation}
Then, using $p_++p_-=1$, $p_-$ can also be obtained.
It is worth mentioning  that the above result reduces to the corresponding result of fixed boundaries  \cite{Firouzjahi:2018vet} in the limit $b\rightarrow0$. 

Next we  study the time average in this case. Again, considering $\phi_+=y$ when the field hits the right boundary, we have
\begin{equation}\label{ubarriertime}
\begin{split}
  \left<\mathcal{N}\right>=& \int_0^b{(\phi_0-\phi_-)(y-\phi_0) \frac{dy}{b}}\\
  =&\left(\phi _0-\phi _-\right) \big(\frac{b}{2}- \phi _0\big) \, .
\end{split}
\end{equation}

\begin{figure}
    \centering
    \begin{subfigure}[t]{0.48\textwidth}
        \centering
        \includegraphics[width=\linewidth]{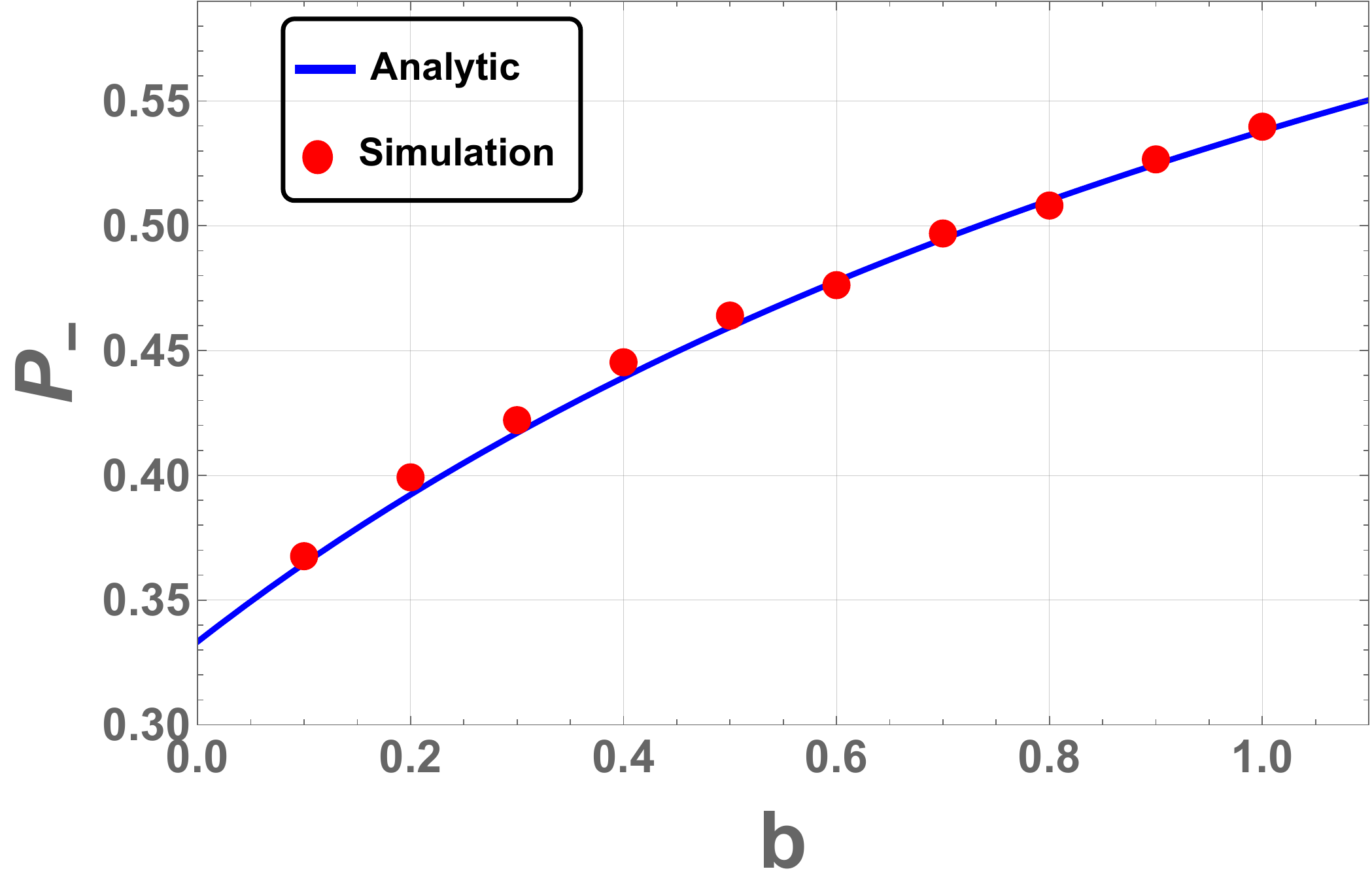} 
        \caption{} \label{uniform-p-b}
    \end{subfigure}
    \hfill
    \begin{subfigure}[t]{0.48\textwidth}
        \centering
        \includegraphics[width=\linewidth]{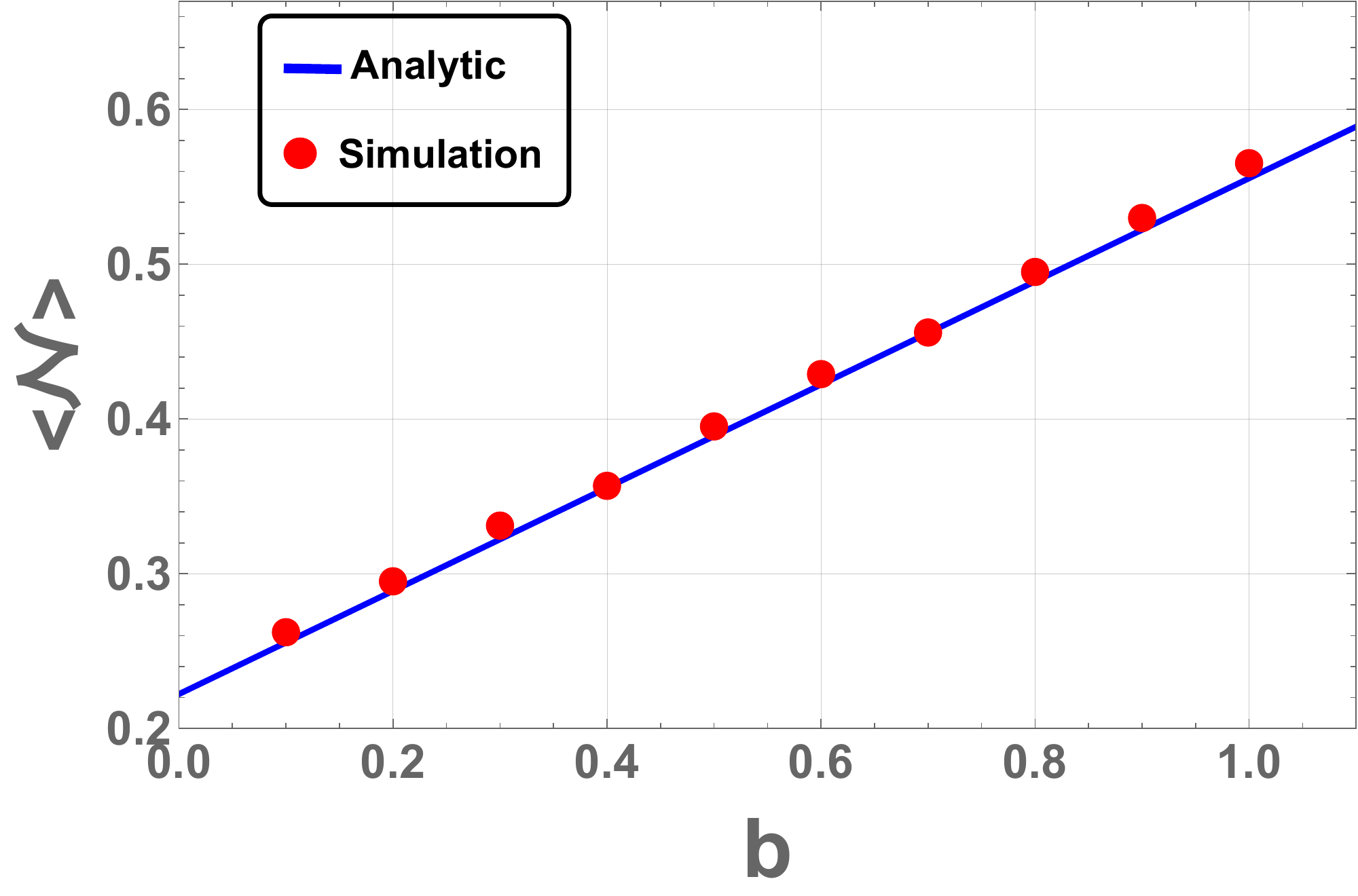} 
        \caption{} \label{uniform-N-b}
    \end{subfigure}
     \caption{The behaviour of $p_-$ (left) and ${\langle{\cal{N}\rangle}}$ (right) versus $b$ for the stochastic boundary with the uniform distribution. Here we have set $\phi_-=-1$ and $\phi_0=-\frac{1}{3}$. As one expects, as $b$  increase  
     both $p_-$ and $\langle \cN \rangle$ increase.   The red points show the data obtained from simulation. The linear (non-linear) dependence of $\langle \cN \rangle$ ($p_-$) on $b$ can be seen in these plots. 
     } 
    %\label{Pm-x0}
      \label{uniform-pN}
\end{figure}

Fig. \ref{uniform-pN} shows the behaviour of $p_-$ and $\left<\mathcal{N}\right>$ with respect to $b$ in the case of uniform distribution.  
As one expects, both $p_-$  
and $\left<\mathcal{N}\right>$ increase a $b$ increases. As seen from our results, 
$\left<\mathcal{N}\right>$  depends linearly on $b$ while $p_-$ depends non-linearly on $b$.   Also in Fig. \ref{uniform-pN} we  present the data for $p_-$ and $\left<\mathcal{N}\right>$ obtained from simulations which are in very good agreements with our analytical results.

%%%%%%%%%%%%%%%%%%%%%%%%%%%%%%%%%%%%%%%%

%%%%%%%%%%%%%%%%%%%%%%%%%%%%%%%%%%%%%%%%%%%%%%%%%%%%%%%%%%%%%%%%%%%%%%%%%%

\section{Summary and Discussions}
\label{summary}

Within the context of stochastic inflation we have studied the Brownian motion of a field which is restricted to move between two boundaries, one of them is fixed at a constant value while the other one undergoes a Brownian motion. There are a number of physical interests to consider this setup. For example in the models of multiple field inflation, there are scenarios where the surface of the end of inflation is modulated by the quantum fluctuations of a light spectator field. There are additional contributions to curvature perturbations from the quantum fluctuations of the spectator field(s). Furthermore, in the landscape of inflation in the UV region of the field space the system may experience a period of eternal inflation in which both the inflaton field and the boundaries may experience  Brownian motion.

We have presented the Langevin equation in various related examples and have calculated the mean number of e-folds $\langle \cN \rangle$ and the first hitting probabilities $p_\pm$ for the field to hit either of the boundaries. 
First, we studied the case in which the classical drift force of the field is dominant compared to the Brownian motion of both the field and the boundary. This setup mimics  models of slow-roll inflation in which the surface of end of inflation is modulated  by a light spectator field.  Using the boundary crossing approach, the average of e-folding number as well as the power spectrum in the drift dominated regime were obtained. It was shown, as expected, that the corrections in power spectrum is proportional to the square of the  amplitude of the noise associated with the boundary.

Our main interest in this work was for the case where the system was diffusion dominated in which the classical drift force of the field 
is subdominant compared to the diffusion forces. We considered two boundaries in which one of them (here  the right boundary) undergoes a Brownian motion in the region $[0,b]$. Solving the corresponding Fokker-Planck equation 
we have obtained the time dependent  probability density functions (PDF) of the  field, $f(\phi, N)$, and the Brownian boundary, $f_+(\phi_+, N)$.  While $f(\phi, N)$ has a Gaussian distribution the solution  for  $f_+(\phi_+, N)$ is given in a series expansion. Equipped with these PDFs and using the Volterra integral equations 
we have calculated the first hitting probabilities $p_\pm$ and the mean number of e-folds $\langle \cN \rangle$ to few orders of $m$ in the series expansion. The result for leading order $p_+^{LO}$ with $m=0$ matches with the case with two fixed boundaries, one at $\phi_-$ and the other at $\phi_+=\frac{b}{2}$. Next we considered 
higher order terms in the series expansion with $m=1$ (NLO), $m=2$ ($N^2LO$) and have looked at the corresponding corrections in $p_\pm$. The series converges to its final result rapidly, specially for smaller value of $b$ as the higher order corrections become exponentially suppressed. As a general conclusion we have shown that  increasing $b$ resulted in higher probability  $p_-$. This is understood easily by noting that by increasing $b$ the right boundary on average moves further away from the field so it is more likely that the field hits the left boundary first. In addition, the behaviour of $p_-$ versus the position of the fixed boundary $\phi_-$ as well as the initial condition of the field $\phi_0$ are studied as well. The figures confirm that, with all other initial conditions held fixed, when $\phi_0$ approaches $\phi_-$ the probability $p_-$ increases while the result is reversed when   $\phi_0$ moves away from $\phi_-$. 

We have calculated $\langle \cN \rangle$ to leading and next to leading orders as well.   As an interesting conclusion of this study, we have compared the result for $\langle \cN \rangle$  to the result in the setup where both boundaries are held fixed. For the  initial condition $\phi_{+}^{(0)}<\frac{b}{2}$ the value of  $\langle \cN \rangle$ in our setup is larger than compared to the case of fixed boundaries. This is because in our setup, the Brownian boundary can explore the classically forbidden region $(\phi_{+}^{(0)}, b)$ as well so, effectively, the length of its journey in field space is larger than the case of the fixed boundaries. On the other hand, for the case with the initial condition  $\phi_{+}^{(0)}>\frac{b}{2}$ the Brownian boundary explores lesser of distances in field space compared to the case of fixed boundary and, as a result, ${\langle{\cal{N}\rangle}}$ reduces compared to the model with the fixed boundaries. Finally, we also studied the effects of $D$, the diffusion amplitude of the jumps of the stochastic boundary, on  $p_\pm$ and  ${\langle{\cal{N}\rangle}}$. Our conclusion is that the roles of $D$ are opposite to the effects of $b$. Specifically, increasing (decreasing) the magnitude of $D$ yields to smaller (larger) values of $p_-$ and  ${\langle{\cal{N}\rangle}}$. 

There are a number of directions in which the current work can be extended. One interesting case to study is the setup where both boundaries undergo 
Brownian motion. In addition, we can assume the boundaries to have different boundary conditions, corresponding to whether the boundary is absorbing or reflecting.   Another interesting example to study 
is when we have more than one stochastic field in the presence of stochastic boundaries. In the context of inflation, this corresponds to the setup  with $N\ge 3$ fields in which $N-1$ fields collectively drive inflation while the remaining  field is a spectator field which modulates the surface of end of inflation generating stochasticity at the surface of end of inflation.  

%%%%%%%%%%%%%%%%%%%%%%%%%%%%%%%%%%%%%%%%%%%%%
\vspace{0.7cm}

 {\bf Acknowledgments:}  H.F. would like to thank YITP, Kyoto University for the hospitality during the workshop ``Gravity: Current challenge in black hole physics and cosmology'' where this work was in its final stage.  H. F. acknowledges partial support from  the ``Saramadan'' federation of Iran. A. N. would like to thank S. Hooshangi for helpful discussions. 

%\vspace{0.5cm}
%%%%%%%%%%%%%%%%%%%%%%%%%%%%%%%%%%%%%%%%%%%%%%%%%%%%%%%%%%%%%%%%%%%%%%%%%%%%%%%%%%
\appendix

%%%%%%%%%%%%%%%%%%%%%%%%%%%%%%%%%%%%%%%%%%%%%%%%%%%%%%%%%%%%%%%%%%%%%%%%%%%%%%%%%%
\section{The proof to Volterra integral equations with the Brownian boundary }\label{proof}

In this section we follow the same process used by \cite{Bou:1990} to obtain a similar set of equations for $\gamma^\pm(\tau)$ in the case where one of the boundaries has Brownian motion.  We denote the two boundaries by $S_1$ and $S_2$ respectively. Let $g(S,t|\phi_0,t_0)$ denote the first passage time PDF to cross a boundary $S$ with $\phi_0$ and $t_0$ given as the initial values for position and time.  We then can write:
\begin{equation}\label{gs1x0}
\begin{split}
    g(S_1,t|\phi_0,t_0)=\gamma^-(t|\phi_0,t_0)+\int_ydy\int^t_{t_0}d\tau\gamma^+(\tau|\phi_0,t_0)g(S_1(t),t|S_2(\tau)=y,\tau)f_{+}(S_2=y,\tau)
    \end{split}
\end{equation}
and
\begin{equation}\label{gs2x0}
\begin{split}
    &g(S_2,t|\phi_0,t_0)=\gamma^+(t|\phi_0,t_0)+\int^t_{t_0}d\tau\gamma^-(\tau|\phi_0,t_0)g(S_2,t|S_1,\tau) \, .
    \end{split}
\end{equation}
 Now we have the following lemma:\\\\
 
 Lemma 1.\\\\
 If $P(x\geq S_2(t)|\phi_0,t_0)$ denotes the transition function of $S_2$ then we can write:
 \begin{equation}
     P(x\geq S_2(t)|\phi_0,t_0)=\int_y P(x\geq S_2(t)=y|\phi_0,t_0,S_2(t)=y)f_{+}(y,t)dy \, .
 \end{equation}
 Proof. We write
 \begin{equation}
 \begin{split}
     P(x\geq S_2(t)|\phi_0,t_0)&=\frac{P(x\geq S_2,\phi_0,y_0)}{P_i(\phi_0,t_0)}=\sum_y\frac{P(x\geq S_2(t),\phi_0,y_0\cap S_2(t)=y)}{P_i(\phi_0,t_0)}\\&=\sum_y\frac{P(x\geq S_2(t)|\phi_0,y_0\cap S_2(t)=y)}{P_i(\phi_0,t_0)}P(\phi_0,t_0,S_2(t)=y)\\&=\sum_y\frac{P(x\geq S_2(t)|\phi_0,y_0\cap S_2(t)=y)}{P_i(\phi_0,t_0)}P_i(\phi_0,t_0)f_{+}(y,t)\\&=\int_y P(x\geq S_2=y|\phi_0,t_0,S_2=y)f_{+}(y)dy.\qquad\qquad\qquad\qquad\qquad\qquad\qquad \quad  \square
     \end{split}
 \end{equation}
Here $P_i$ denotes the PDF that the initial condition is fixed at $\phi_0$ and $t_0$ and in the last line we have used the fact that the PDF of initial condition is independent of $f_{+}$.
 Another similar lemma may be expressed as follows:
 
Lemma 2.\\\\ For any $x\notin(S_1,S_2)$ one has
\begin{equation}
\begin{split}\label{fgamma}
    f(x,t|\phi_0,t_0)&=\int^t_{t_0}d\tau\Big(\gamma^-(\tau|\phi_0,t_0)f(x,t|S_1,\tau)+\\&\int_ydy\gamma^+(\tau|\phi_0,t_0)f(x,t|S_2=y,\tau)f_{+}(S_2=y,\tau)\Big) \, .
    \end{split}
\end{equation}

Proof.
If $x\geq S_2$ then we have:
\begin{equation}\label{fing}
f(x,t|\phi_0,t_0)=\int^t_{t_0}g(S_2(\tau),\tau|\phi_0,t_0)f(x,t|S_2(\tau),\tau)d\tau \, .
\end{equation}
The same as what we had in  lemma 1, one can show that 
\begin{equation}\label{fg}
   g(S_2(\tau),\tau|\phi_0,t_0)=\int_y g(S_2(\tau),\tau|\phi_0,t_0,S_2(\tau)=y)f_{+}(y,\tau)dy \, ,
\end{equation}
and so we write
\begin{equation}
\begin{split}
f(x,t|\phi_0,t_0)=\int^t_{t_0}\int_yg(S_2(\tau),\tau|\phi_0,t_0,S_2(\tau)=y)f(x,t|S_2(\tau)=y,\tau)f_{+}(y,\tau)dy d\tau \, .
\end{split}
\end{equation}
Then using Eqs. \eqref{gs1x0} and \eqref{gs2x0} we obtain
\begin{equation}
\begin{split}
    f(x,t|\phi_0,t_0)&=\int\int dyd\tau\gamma^+(\tau|\phi_0,t_0)f(x,t|S_2(\tau)=y,\tau)f_{+}(y,\tau)+\\\int d\rho \gamma^-(\rho|\phi_0,t_0)&\int d\tau\int_y dy g(S_2(\tau)=y,\tau|S_1(\rho),\rho,S_2(\tau)=y) f(x,t|S_2(\tau)=y,\tau) f_{+}(y,\tau) .
    \end{split}
\end{equation}
Using Eq. \eqref{fing} and replacing $\phi_0=S_1(\tau)$ and $t_0=\rho$ we obtain the result. The proof for $x\leq S_1$ is similar.\qquad\qquad\qquad\qquad\qquad\qquad\qquad \qquad \qquad \qquad \qquad \qquad  \qquad \qquad \qquad \qquad \qquad$ \square$

Now we are ready for the following lemma which gives a proof to Eqs. (\ref{gammaplus}) and (\ref{gammaminus}).

Lemma 3.\\

Let's define
\begin{equation}
\begin{split}
        &\psi(x\geq S_1(t),t|y,\tau)\equiv \frac{d}{dt}F(S_1(t),t|y,\tau),\\
    &\psi(x\geq S_2(t),t|y,\tau)\equiv\\& \frac{d}{dt}\Big(\int_xF(S_2(t)=x,t|y,\tau,S_2(\tau)=y,S_2(t)=x)f_{+}(y,t)f_{+}(x,t|y,\tau)\Big),
    \end{split}
\end{equation}
then we have
\begin{equation}\label{gammaminusfinal}
\begin{split}
    \gamma^-(t|\phi_0,t_0)&=2\psi(S_1,t|\phi_0,t_0)-2\int^t_{t_0}d\tau\Big(\gamma^-(\tau|\phi_0,t_0)\psi(S_1,t|S_1,\tau)\\&+\int_ydy\gamma^+(\tau|\phi_0,t_0)\psi(S_1,t|S_2=y,\tau)f_{+}(y,\tau)\Big),
    \end{split}
\end{equation}
and 
\begin{equation}\label{gammaplusfinal}
\begin{split}
    \gamma^+(t|\phi_0,t_0)&=-2\psi(S_2,t|\phi_0,t_0)+\\&2\int^t_{t_0}d\tau\Big(\int_x\int_y\gamma^+(\tau|\phi_0,t_0)\psi(S_2(t),t|S_2(\tau)=y,\tau)f_{+}(y,\tau)dxdy\\&+\int_ydy\gamma^+(\tau|\phi_0,t_0)\psi(S_2(t)=y,t|S_1,\tau)f_{+}(y,\tau)\Big)
    \end{split}
\end{equation}
\\Proof:

We prove the second equation and the first one can be proved similarly. To this end we firstly note that
\begin{equation}
    F(S_2(t),t|S_2(\tau)=y,\tau)=\int_xF(S_2(t)=x,t|S_2(\tau)=y,\tau)f_{+}(x,t|y,\tau)dx.
\end{equation}
To prove this relation we write:
\begin{equation}\label{sipp}
\begin{split}
 &F(S_2(t),t|S_2(\tau)=y,\tau)=P(\phi\leq S_2(t),t|S_2(\tau)=y,\tau)\\\\&=\frac{P(\phi\leq S_2(t)\cap S_2(\tau)=y\cap\text{The field starts at}\quad S_2(\tau))}{P(S_2(\tau)=y\cap\text{The field starts at} S_2(\tau)=y)}\\\\&=\sum_x\frac{P(\phi\leq S_2(t)\cap \alpha \cap S_2(t)=x)}{P(\alpha)}=\sum_x\frac{P(\phi\leq S_2(t)| \alpha , S_2(t)=x)}{P(\alpha)}P(\alpha , S_2(t)=x),
 \end{split}
\end{equation}
where in the second equality we have defined the event in the denominator by $\alpha$. 

Now using the Bayes' theorem we can write:
\begin{equation}\label{bayse}
   \frac{P(\alpha,S_2(t)=x)}{P(\alpha)}=P(S_2(t)=x|\alpha) \, .
\end{equation}
As the the field evolves independently of $S_2$ then one can write Eq. \eqref{bayse} as
\begin{equation}
   P(S_2(t)=x|\alpha)=P(S_2(t)=x|S_2(\tau)=y)\equiv f_{+}(x,t|y,\tau).
\end{equation}
By writing the last summation as an integral then we can write:
\begin{equation}\label{Fpp}
F(S_2(t),t|S_2(\tau)=y,\tau)=\int_x F(S_2(t)=x|S_2(\tau)=y)f_{+}(x,t|y,\tau)dx. 
\end{equation}
Now we are at a stage to prove the second equation. Let $x\geq S_2(t)$. Then by integrating Eq. \eqref{fgamma} between a constant boundary $r_2>S_2(t)$ and $S_2(t)$ and defining $F_c(S_2,t|\phi_0,t_0)=1-F(S_2,t|\phi_0,t_0)$ we obtain
\begin{equation}\label{fceq}
\begin{split}
    &F_c(S_2,t|\phi_0,t_0)=\\&\int d\tau(\gamma^-(\tau|\phi_0,t_0)F_c(S_2,t|S_1,\tau)+\gamma^+(\tau|\phi_0,t_0)\int_yF_c(S_2(t)|S_2(\tau)=y,\tau)f_{+}(y,\tau)dy).
    \end{split}
\end{equation}
By taking the derivative of Eq. \eqref{fceq} with respect to time and using the following relations
\begin{equation}
\begin{split}
    &\lim_{\tau\rightarrow t}F(S_1(t)|S_2(t))=0,\\&
    \lim_{\tau\rightarrow t}F(S_1(t)|S_2(\tau))=\frac{1}{2},\\&
     \lim_{\tau\rightarrow t}F(S_2(t)|S_1(\tau))=1\\&
     \lim_{\tau\rightarrow t}F(S_2(t)|S_2(\tau))=\frac{1}{2}.
    \end{split}
\end{equation}
 as well as Eq. \eqref{Fpp} one obtains Eq. \eqref{gammaplusfinal}. Eq. \eqref{gammaminusfinal} can be obtained in a similar manner.

At last it can be useful to prove that the distribution of the stochastic boundary at the  crossing time the same as the distribution of the boundary itself without considering any stopping time. In other words the distribution of the boundary is independent of the first crossing time distribution. To this end we define the $n$-th moment of the field at time $t$ with the event A while the event that the field crosses the stochastic boundary at time $t$ is described with event B ($\text{B}^c$ represents the complement of B). Thus one can write 
\ba
    E(\text{A})=E(\text{A}|\text{B}) P(\text{B}) +E(\text{A}|\text{B}^c) P(\text{B}^c) \, .
\ea
As $A$ is obviously independent of $\text{B}^c$ and using the fact that $P(\text{B})+P(\text{B}^c)=1$, one can easily show that $ E(\text{A})=E(\text{A}|\text{B})$. Therefore all moments of $A$ are equal to all moments of conditional $A$, so one can deduce that their distributions are the same as well.

\section{First crossing probabilities }
\label{prob}

In this Appendix we find a solution for $\gamma^\pm(t)$. To this end we suppose that the Brownian boundary has a small drift changing linearly with time. In other words we suppose that the 
\begin{equation}
\label{driftedlangevin}
   \phi_{+}(N)=vN+D W_+(N).
\end{equation}
Assuming the boundary has a small drift we should modify $f_{+}(x,N)$ 
where we assume that $v$ is so small that for a significant time interval we have 
\begin{equation}
    |v N|\ll D.
\end{equation}
 If $F_{+d}(x,N)$ denotes the transition PDF of the boundary with drift then one can write:
\begin{equation}
    F_{+d}(x,N)=P(\phi_{+}+v N\leq x,N)=P(\phi_{+}\leq x-vN,N).
\end{equation}
Then the probability density of the Brownian boundary, $f_{+d}(\phi_+, N)$,  is simply given by:
\begin{equation}\label{densitydrift}
    f_{+d}(x,N)=\frac{\partial F_{+d}(x,N)}{\partial x}=\frac{\partial P(\phi_+ \leq x-vN,N)}{\partial x}=f_{+}(x-v N,N).
\end{equation}
One can easily check that the above result satisfies the following Fokker-Planck equation in the presence of the drift:
   \begin{equation}
   \label{fokkerbarrier}
    \frac{\partial f_{+d}(x,N)}{\partial N}=-v\frac{\partial f_{+d}(x,N)}{\partial x}+\frac{D^2}{2}\frac{\partial^2 f_{+d}(x,N)}{\partial x^2}
\end{equation}
with the boundary conditions the same as Eqs. \eqref{BC} and \eqref{delta}. From now on, using Eq. (\ref{fp}), we can obtain $f_{+}(x-v N,N)$ and we will set $v=0$ for a boundary with pure Brownian motion.

Using Eqs. (\ref{densitydrift}), (\ref{gammaminusfinal}) and (\ref{gammaplusfinal}) we obtain
\begin{equation}\label{gamapluseq}
\begin{split}
&\Gamma^+(s)+\frac{e^{\sqrt{2s} b }-1}{\sqrt{2s} b } \left[\Gamma^-(s)-e^{\sqrt{2s}  (\phi_0-b)}\right]e^{\sqrt{2s}  (\phi_--b)}\\&-4\sum_{m=1}^\infty\Bigg(\Bigg[\frac{  b^2 s\left((-1)^m-e^{ \sqrt{2b^2 s+\pi ^2 m^2 D^2}}\right)}{ \left(2b^2 s+\pi ^2 m^2 (D^2+1)\right)\sqrt{2b^2 s+\pi ^2 m^2 D^2}}\Bigg]\Gamma^-\left(\frac{m^2\pi ^2 D^2 }{2b^2}+s\right) \\&e^{\frac{ (\phi_--b)}{b} \sqrt{2b^2 s+\pi ^2 m^2 D^2}}\cos \left(\frac{  m\pi }{b}\phi_{+}^{(0)}\right)\Bigg)+4\sum_{n=1}^\infty\Bigg(\Bigg[\frac{ b^2 s \left((-1)^n-e^{ \sqrt{2b^2 s+\pi ^2 n^2 D^2}}\right)}{ \left(2b^2 s+ \pi ^2 n^2 (D^2+1)\right)\sqrt{2b^2 s+\pi ^2 n^2 D^2}}\Bigg]\\&e^{\frac{ \left(\phi _0-b\right)}{b} \sqrt{2b^2 s+\pi ^2 n^2 D^2}}\cos \left(\frac{  n\pi }{b}\phi_{+}^{(0)}\right)\Bigg)+vQ^+=0,
\end{split}
\end{equation}
and
\begin{equation}\label{gamaminuseq}
\begin{split}
    &\Gamma^-(s)-e^{\sqrt{2s}  \left(\phi_--\phi _0\right)}+\frac{\left(e^{  \sqrt{2s}b}-1\right)}{\sqrt{2s} b } e^{\sqrt{2s} \left(\phi_--b\right)}\,\Gamma^+(s)\\&
    -2\sqrt{2s}\sum_{m=1}^\infty\Bigg[\frac{ b  \left((-1)^m-e^{\sqrt{2s} b }\right)  }{2 b^2 s+ m^2\pi ^2}\Bigg]e^{\sqrt{2s}  (\phi_--b)}\Gamma^+\left(\frac{ m^2\pi ^2 D^2}{2b^2}+s\right)\cos \left(\frac{m\pi }{b}\phi_{+}^{(0)}\right) +vQ^-=0 \, .
    \end{split}
\end{equation}
In the above equations we have expanded the equations up to first order of $v$ while $Q^\pm$ are some functions depending on our initial values and the Laplace parameter that we have not presented their explicit forms here as they involve large expressions. The above equations are used to obtain Eqs. (\ref{pplus}) and (\ref{N-NLO}).

\end{document}